\documentclass[twocolumn]{aastex63}

\usepackage{epsfig,natbib}
\usepackage{graphicx}
\usepackage{amsmath}
\usepackage{booktabs}
\usepackage{longtable}
\usepackage{xspace}
\usepackage{hyperref}
\usepackage[T1]{fontenc}
\usepackage{lipsum}
\shortauthors{MacDougall et al.}
\shorttitle{TOI-1272}
\begin{document}

\title{The TESS-Keck Survey. XIII. An Eccentric Hot Neptune with a Similar-Mass Outer Companion around TOI-1272}

\author[0000-0003-2562-9043]{Mason G.\ MacDougall}
\affiliation{Department of Physics \& Astronomy, University of California Los Angeles, Los Angeles, CA 90095, USA}

\author[0000-0003-0967-2893]{Erik A.\ Petigura}
\affiliation{Department of Physics \& Astronomy, University of California Los Angeles, Los Angeles, CA 90095, USA}

\author[0000-0002-3551-279X]{Tara Fetherolf}
\affiliation{Department of Earth and Planetary Sciences, University of California Riverside, Riverside, CA 92521, USA}

\author[0000-0001-7708-2364]{Corey Beard}
\affiliation{Department of Physics \& Astronomy, University of California Irvine, Irvine, CA 92697, USA}

\author[0000-0001-8342-7736]{Jack Lubin}
\affiliation{Department of Physics \& Astronomy, University of California Irvine, Irvine, CA 92697, USA}


\author[0000-0002-9751-2664]{Isabel Angelo}
\affiliation{Department of Physics \& Astronomy, University of California Los Angeles, Los Angeles, CA 90095, USA}

\author[0000-0002-7030-9519]{Natalie M. Batalha}
\affiliation{Department of Astronomy and Astrophysics, University of California Santa Cruz, Santa Cruz, CA 95060, USA}

\author[0000-0003-0012-9093]{Aida Behmard}
\altaffiliation{NSF Graduate Research Fellow}
\affiliation{Division of Geological and Planetary Science, California Institute of Technology, Pasadena, CA 91125, USA}

\author[0000-0002-3199-2888]{Sarah Blunt}
\altaffiliation{NSF Graduate Research Fellow}
\affiliation{Department of Astronomy, California Institute of Technology, Pasadena, CA 91125, USA}

\author[0000-0002-4480-310X]{Casey Brinkman}
\altaffiliation{NSF Graduate Research Fellow}
\affiliation{Institute for Astronomy, University of Hawai`i, Honolulu, HI 96822, USA}

\author[0000-0003-1125-2564]{Ashley Chontos}
\altaffiliation{NSF Graduate Research Fellow}
\affiliation{Institute for Astronomy, University of Hawai`i, Honolulu, HI 96822, USA}

\author[0000-0002-1835-1891]{Ian J. M. Crossfield}
\affiliation{Department of Physics \& Astronomy, University of Kansas, Lawrence, KS 66045, USA}

\author[0000-0002-8958-0683]{Fei Dai}
\affiliation{Division of Geological and Planetary Science, California Institute of Technology, Pasadena, CA 91125, USA}

\author[0000-0002-4297-5506]{Paul A. Dalba}
\altaffiliation{NSF Astronomy and Astrophysics Postdoctoral Fellow}
\affiliation{Department of Astronomy and Astrophysics, University of California Santa Cruz, Santa Cruz, CA 95060, USA}
\affiliation{Department of Earth and Planetary Sciences, University of California, Riverside, CA 92521, USA}

\author[0000-0001-8189-0233]{Courtney Dressing}
\affiliation{Department of Astronomy, University of California Berkeley, Berkeley, CA 94720, USA}

\author[0000-0003-3504-5316]{Benjamin Fulton}
\affiliation{NASA Exoplanet Science Institute/Caltech-IPAC, Pasadena, CA 91125, USA}

\author[0000-0002-8965-3969]{Steven Giacalone}
\affiliation{Department of Astronomy, University of California Berkeley, Berkeley, CA 94720, USA}

\author[0000-0002-0139-4756]{Michelle L. Hill}
\affiliation{Department of Earth and Planetary Sciences, University of California Riverside, Riverside, CA 92521, USA}

\author[0000-0001-8638-0320]{Andrew W.\ Howard}
\affiliation{Department of Astronomy, California Institute of Technology, Pasadena, CA 91125, USA}

\author[0000-0001-8832-4488]{Daniel Huber}
\affiliation{Institute for Astronomy, University of Hawai`i, Honolulu, HI 96822, USA}

\author[0000-0002-0531-1073]{Howard Isaacson}
\affiliation{Department of Astronomy, University of California Berkeley, Berkeley, CA 94720, USA}
\affiliation{Centre for Astrophysics, University of Southern Queensland, Toowoomba, QLD, Australia}

\author[0000-0002-7084-0529]{Stephen R. Kane}
\affiliation{Department of Earth and Planetary Sciences, University of California Riverside, Riverside, CA 92521, USA}

\author{Molly Kosiarek}
\affiliation{Department of Astronomy and Astrophysics, University of California Santa Cruz, Santa Cruz, CA 95060, USA}

\author{Andrew Mayo}
\affiliation{Department of Astronomy, University of California Berkeley, Berkeley, CA 94720, USA}

\author[0000-0003-4603-556X]{Teo Mo\v{c}nik}
\affiliation{Gemini Observatory/NSF's NOIRLab, 670 N. A'ohoku Place, Hilo, HI 96720, USA}

\author[0000-0001-8898-8284]{Joseph M. Akana Murphy}
\altaffiliation{NSF Graduate Research Fellow}
\affiliation{Department of Astronomy and Astrophysics, University of California Santa Cruz, Santa Cruz, CA 95060, USA}

\author{Daria Pidhorodetska}
\affiliation{Department of Earth and Planetary Sciences, University of California Riverside, Riverside, CA 92521, USA}

\author[0000-0001-7047-8681]{Alex Polanski}
\affiliation{Department of Physics \& Astronomy, University of Kansas, Lawrence, KS 66045, USA}

\author{Malena Rice}
\altaffiliation{NSF Graduate Research Fellow}
\affiliation{Department of Astronomy, Yale University, New Haven, CT 06520, USA}

\author[0000-0003-0149-9678]{Paul Robertson}
\affiliation{Department of Physics \& Astronomy, University of California Irvine, Irvine, CA 92697, USA}

\author[0000-0001-8391-5182]{Lee J.\ Rosenthal}
\affiliation{Department of Astronomy, California Institute of Technology, Pasadena, CA 91125, USA}

\author[0000-0001-8127-5775]{Arpita Roy}
\affiliation{Space Telescope Science Institute, Baltimore, MD 21218, USA}
\affiliation{Department of Physics and Astronomy, Johns Hopkins University, Baltimore, MD 21218, USA}

\author[0000-0003-3856-3143]{Ryan A. Rubenzahl}
\altaffiliation{NSF Graduate Research Fellow}
\affiliation{Department of Astronomy, California Institute of Technology, Pasadena, CA 91125, USA}

\author[0000-0003-3623-7280]{Nicholas Scarsdale}
\affiliation{Department of Astronomy and Astrophysics, University of California Santa Cruz, Santa Cruz, CA 95060, USA}

\author[0000-0002-1845-2617]{Emma V. Turtelboom}
\affiliation{Department of Astronomy, University of California Berkeley, Berkeley, CA 94720, USA}

\author{Dakotah Tyler}
\affiliation{Department of Physics \& Astronomy, University of California Los Angeles, Los Angeles, CA 90095, USA}

\author[0000-0002-4290-6826]{Judah Van Zandt}
\affiliation{Department of Physics \& Astronomy, University of California Los Angeles, Los Angeles, CA 90095, USA}

\author[0000-0002-3725-3058]{Lauren M. Weiss}
\affiliation{Department of Physics, University of Notre Dame, Notre Dame, IN 46556, USA}


\author[0000-0002-2341-3233]{Emma Esparza-Borges}
\affiliation{Instituto de Astrof\'{i}sica de Canarias (IAC), 38205 La Laguna, Tenerife, Spain}
\affiliation{Departamento de Astrofísica, Universidad de La Laguna, 38206 La Laguna, Tenerife, Spain}

\author[0000-0002-4909-5763]{Akihiko Fukui}
\affiliation{Komaba Institute for Science, The University of Tokyo, 3-8-1 Komaba, Meguro, Tokyo 153-8902, Japan}
\affiliation{Instituto de Astrof\'{i}sica de Canarias (IAC), 38205 La Laguna, Tenerife, Spain}

\author[0000-0002-6480-3799]{Keisuke Isogai}
\affil{Okayama Observatory, Kyoto University, 3037-5 Honjo, Kamogatacho, Asakuchi, Okayama 719-0232, Japan}
\affil{Department of Multi-Disciplinary Sciences, Graduate School of Arts and Sciences, The University of Tokyo, 3-8-1 Komaba, Meguro, Tokyo 153-8902, Japan}

\author[0000-0003-1205-5108]{Kiyoe Kawauchi}
\affiliation{Instituto de Astrof\'{i}sica de Canarias (IAC), 38205 La Laguna, Tenerife, Spain}

\author[0000-0003-1368-6593]{Mayuko Mori}
\affil{Department of Astronomy, Graduate School of Science, The University of Tokyo, 7-3-1 Hongo, Bunkyo-ku, Tokyo 113-0033, Japan}

\author[0000-0001-9087-1245]{Felipe Murgas}
\affiliation{Instituto de Astrof\'{i}sica de Canarias (IAC), 38205 La Laguna, Tenerife, Spain}
\affiliation{Departamento de Astrofísica, Universidad de La Laguna, 38206 La Laguna, Tenerife, Spain}

\author[0000-0001-8511-2981]{Norio Narita}
\affiliation{Komaba Institute for Science, The University of Tokyo, 3-8-1 Komaba, Meguro, Tokyo 153-8902, Japan}
\affiliation{Instituto de Astrof\'{i}sica de Canarias (IAC), 38205 La Laguna, Tenerife, Spain}

\author[0000-0003-1510-8981]{Taku Nishiumi}
\affil{Department of Astronomical Science, The Graduated University for Advanced Studies, SOKENDAI, 2-21-1, Osawa, Mitaka, Tokyo, 181-8588, Japan}
\affil{Astrobiology Center, 2-21-1 Osawa, Mitaka, Tokyo 181-8588, Japan}
\affil{Department of Multi-Disciplinary Sciences, Graduate School of Arts and Sciences, The University of Tokyo, 3-8-1 Komaba, Meguro, Tokyo 153-8902, Japan}

\author[0000-0003-0987-1593]{Enric Palle}
\affiliation{Instituto de Astrof\'{i}sica de Canarias (IAC), 38205 La Laguna, Tenerife, Spain}
\affiliation{Departamento de Astrofísica, Universidad de La Laguna, 38206 La Laguna, Tenerife, Spain}

\author[0000-0001-5519-1391]{Hannu Parviainen}
\affiliation{IInstituto de Astrof\'{i}sica de Canarias (IAC), 38205 La Laguna, Tenerife, Spain}
\affiliation{Departamento de Astrofísica, Universidad de La Laguna, 38206 La Laguna, Tenerife, Spain}

\author[0000-0002-7522-8195]{Noriharu Watanabe}
\affil{Department of Multi-Disciplinary Sciences, Graduate School of Arts and Sciences, The University of Tokyo, 3-8-1 Komaba, Meguro, Tokyo 153-8902, Japan}


\author[0000-0002-4715-9460]{Jon M.\ Jenkins}
\affiliation{NASA Ames Research Center, Moffett Field, CA 94035-0001, USA}

\author[0000-0001-9911-7388]{David W.\ Latham}
\affil{Center for Astrophysics | Harvard \& Smithsonian, Cambridge, MA 02138, USA}

\author[0000-0003-2058-6662]{George R.\ Ricker}
\affil{Department of Physics and Kavli Institute for Astrophysics and Space Research, Massachusetts Institute of Technology, Cambridge, MA 02139, USA}

\author[0000-0002-6892-6948]{S. Seager}
\affil{Department of Earth, Atmospheric, and Planetary Sciences, Massachusetts Institute of Technology, Cambridge, MA 02139, USA}
\affil{Department of Physics and Kavli Institute for Astrophysics and Space Research, Massachusetts Institute of Technology, Cambridge, MA 02139, USA}
\affil{Department of Aeronautics and Astronautics, Massachusetts Institute of Technology, Cambridge, MA 02139, USA}

\author[0000-0001-6763-6562]{Roland K.\ Vanderspek}
\affiliation{Department of Physics and Kavli Institute for Astrophysics and Space Research, Massachusetts Institute of Technology, Cambridge, MA 02139, USA}

\author[0000-0002-4265-047X]{Joshua N.\ Winn}
\affiliation{Department of Astrophysical Sciences, Princeton University, Princeton, NJ 08544, USA}


\author[0000-0001-6637-5401]{Allyson~Bieryla}
\affiliation{Center for Astrophysics | Harvard \& Smithsonian, Cambridge, MA 02138, USA}

\author[0000-0003-1963-9616]{Douglas A.\ Caldwell}
\affiliation{NASA Ames Research Center, Moffett Field, CA 94035-0001, USA}

\author[0000-0003-2313-467X]{Diana Dragomir}
\affiliation{Department of Physics and Astronomy, University of New Mexico, Albuquerque, NM 87131, USA}

\author[0000-0002-9113-7162]{M.~M.~Fausnaugh}
\affiliation{Department of Physics and Kavli Institute for Astrophysics and Space Research, Massachusetts Institute of Technology, Cambridge, MA 02139, USA}

\author[0000-0002-4510-2268]{Ismael Mireles}
\affiliation{Department of Physics and Astronomy, University of New Mexico, Albuquerque, NM 87131, USA}

\author[0000-0003-1286-5231]{David R.\ Rodriguez}
\affiliation{Space Telescope Science Institute, Baltimore, MD 21218, USA}


\accepted{24 June 2022}

\begin{abstract}

We report the discovery of an eccentric hot Neptune and a non-transiting outer planet around TOI-1272. We identified the eccentricity of the inner planet, with an orbital period of 3.3 d and $R_{\rm p,b} = 4.1 \pm 0.2$ $R_\earth$, based on a mismatch between the observed transit duration and the expected duration for a circular orbit. Using ground-based radial velocity measurements from the HIRES instrument at the Keck Observatory, we measured the mass of TOI-1272b to be $M_{\rm p,b} = 25 \pm 2$ $M_\earth$. We also confirmed a high eccentricity of $e_b = 0.34 \pm 0.06$, placing TOI-1272b among the most eccentric well-characterized sub-Jovians. We used these RV measurements to also identify a non-transiting outer companion on an 8.7-d orbit with a similar mass of $M_{\rm p,c}$ sin$i= 27 \pm 3$ $M_\earth$ and $e_c \lesssim 0.35$. Dynamically stable planet-planet interactions have likely allowed TOI-1272b to avoid tidal eccentricity decay despite the short circularization timescale expected for a close-in eccentric Neptune. TOI-1272b also maintains an envelope mass fraction of $f_{\rm env} \approx 11\%$ despite its high equilibrium temperature, implying that it may currently be undergoing photoevaporation. This planet joins a small population of short-period Neptune-like planets within the "Hot Neptune Desert" with a poorly understood formation pathway.

\end{abstract}

\section{Introduction}
\label{sec:intro}

The Solar System consists of eight planets from three size categories: terrestrial (0.4--1.0 $R_{\rm \earth}$), ice giant (3.9--4.0 $R_{\rm \earth}$), and gas giant (9.5--11.2 $R_{\rm \earth}$). They are spread out across a 30 AU radial expanse and, with the exception of Mercury, their orbits are nearly circular. All three of these Solar System based patterns (distributions of planet sizes, orbit spacing, and orbital eccentricity) are contradicted by known exoplanetary systems. The prime \emph{Kepler} mission (\citealt{Borucki10a}) revealed that the most common exoplanet sizes are between Earth and Neptune (1.0--3.9 $R_{\rm \earth}$; super-Earths / mini-Neptunes) in addition to a significant population with sizes between Uranus and Saturn (4.0--9.4 $R_{\rm \earth}$; super-Neptunes or sub-Saturns). Moreover, planetary orbits interior to Mercury's orbit are common, as well as orbits with eccentricities of $e > 0.1$ ($>60\%$ of known planets, \citealt{ps}). Such broad demographics demonstrate a variety of possible outcomes to planet formation.

However, there are certain planet characteristics that are less common, even with observational biases accounted for. These include the dearth of extremely high eccentricity planets at short orbital periods, the "gap" in planet radius between super-Earths and mini-Neptunes at $P < 100$ days (\citealt{Fulton17}), and the paucity of short-period Neptunian planets (\citealt{Mazeh16}). The latter was first proposed as a natural consequence of photo-evaporation by planet characterization studies (\citealt{Lopez13}; \citealt{Owen13}). The resulting “Hot Neptune desert” implies that for intermediate size planets between $\sim$10--100 $M_{\rm \earth}$ and $\sim$2--6 $R_{\rm \earth}$, an inefficient formation pathway or an efficient mass loss mechanism sharply differentiates Neptunes from Jupiters at periods of $\lesssim 5$ days. \cite{Owen18} proposed that the upper and lower boundaries of this sparse region of $M_p$-$P$ and $R_p$-$P$ parameter space can be mostly explained by limitations from either eccentricity decay of larger planets or photo-evaporation of smaller planets. The handful of observations of atmosphere-stripped Neptunian cores in the desert further supports this hypothesis (e.g. TOI-849b; \citealt{Armstrong20}), but many questions remain surrounding the formation pathways of such planets.

Eccentricity further complicates the long-term evolutionary history of sub-Jovians on compact orbits. To date, only 8 planets with sizes between 2.0--6.0 $R_{\rm \earth}$ and $M_p < 100$ $M_{\rm \earth}$ have been found to have well-constrained eccentricities of $e > 0.2$, the greatest outlier being Kepler-1656b at $e \approx 0.84 \pm 0.01$ (\citealt{Brady18}). Only 2 of these planets, however, have orbital periods of $P < 5$ days. Hot Jupiter-size planets tend to have longer tidal circularization time-scales and more massive cores that can retain their H/He envelopes during close-in periastron passage, but hot sub-Jovians are more susceptible to eccentricity decay and atmospheric loss. Consequently, the population of hot, eccentric Neptunes with $\gtrsim$10\% H/He envelope mass fraction is small, consisting only of a handful of planets including HAT-P-11b (\citealt{Yee18}) and GJ 436b (\citealt{Lanotte14}). 

In this paper, we discuss TOI-1272b, the latest Neptune to join the sparse population of hot, eccentric sub-Jovians. Leveraging the "photo-eccentric" methodology outlined by \cite{Dawson12} and \cite{Kipping12}, we identified TOI-1272b as a candidate for high eccentricity based on a mismatch between the observed transit duration and the expected duration for a circular orbit. We used this technique as a pre-filter to vet for high-eccentricity candidates based on photometry alone, motivating follow-up radial velocity observations. Similar photometric modeling methods have been applied to \emph{Kepler} target samples (\citealt{Kane12}; \citealt{VE15}; \citealt{Xie16}; \citealt{VE19}), but those studies were not followed up by radial velocity campaigns. We present TOI-1272 as the second system from our photo-eccentric pre-filter study of TOIs, in association with the planetary demographics work being carried out by the TESS-Keck Survey collaboration (TKS; \citealt{Chontos21}).
 
We introduce the TOI-1272 system and discuss the transit profile modeling that we used to identify TOI-1272b as an eccentric planet candidate from photometry (\S\ref{sec:ecc-candidate}). We also describe our follow-up radial velocity observations (\S\ref{sec:rvs}) and analyze our spectroscopic measurements to characterize the properties of the host star, including stellar variability and age (\S\ref{sec:stellar}). From our dense RV data set, we confirm the high eccentricity of TOI-1272b and detect the presence of a non-transiting outer planet (\S\ref{sec:keplerian-model}). Finally, we explore the long-term stability of this system through various dynamical criteria which we use to further constrain our eccentricity measurements (\S\ref{sec:dynamics}). We also place this system in context within the Hot Neptune Desert (\S\ref{sec:context}) and consider possible formation and evolution pathways for TOI-1272b and other hot Neptunes. 

\section{TOI-1272\lowercase{b}: A High-Eccentricity Candidate}
\label{sec:ecc-candidate}

\subsection{Photometry}
\label{sec:photometry}

TOI-1272 was observed by TESS with 2-min-cadence photometry in sectors 15, 16, and 22 between UT 2019 October 10 and 2020 May 11. The time-series photometry was processed by the TESS Science Processing Operations Center pipeline (SPOC; \citealt{Jenkins16}), which first detected the periodic transit signal of TOI-1272b with a wavelet-based, noise-compensating matched filter (\citealt{Jenkins02}; \citealt{Jenkins10}). An initial limb-darkened transit model fit was performed (\citealt{Li19}) and the signature passed a suite of diagnostic tests described by \citealt{Twicken18}, leading this target to be selected as a TOI. 

\begin{figure}[ht]
\centering
\includegraphics[width=0.45\textwidth]{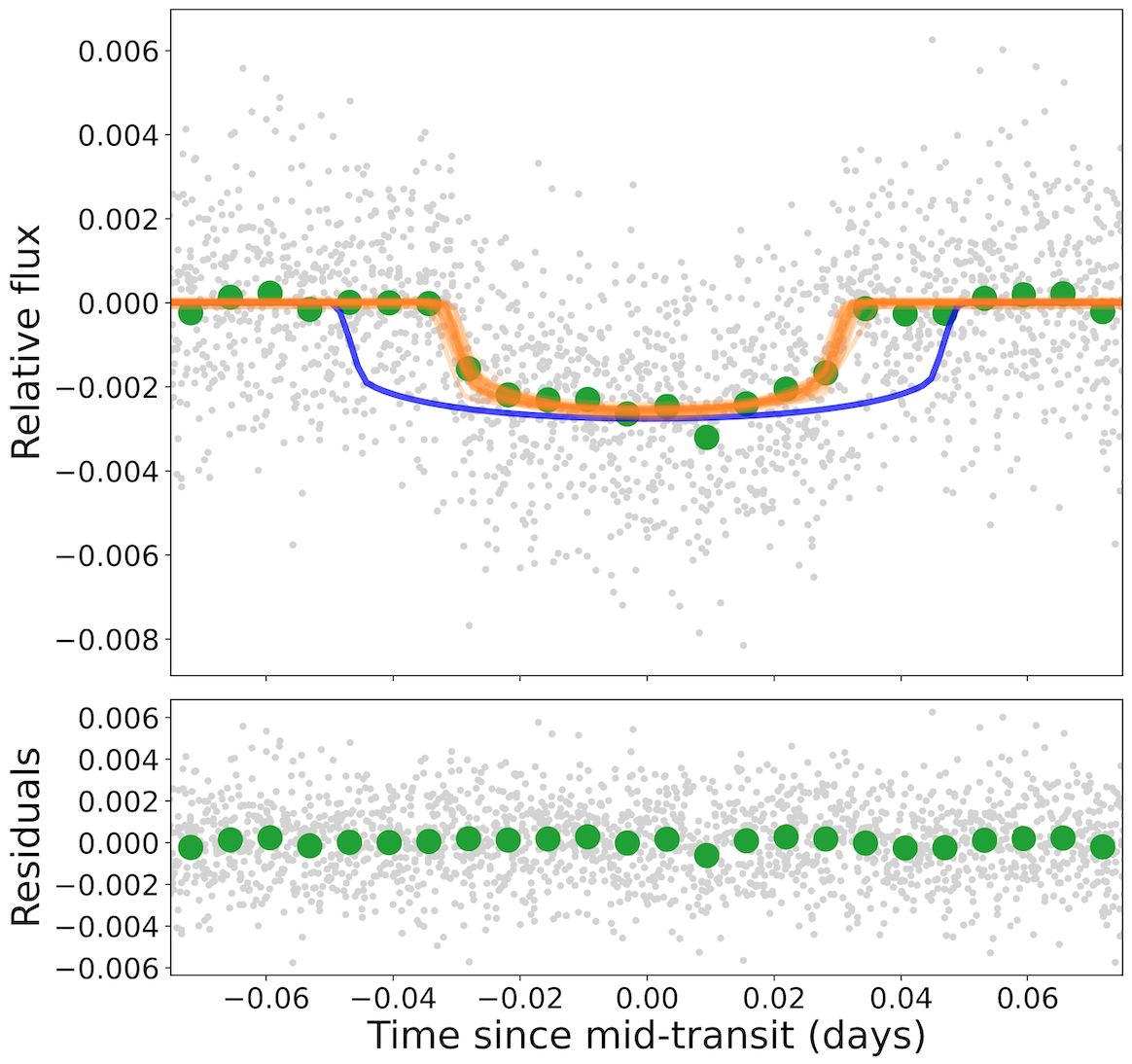}
\caption{Top: Transit models drawn from parameter posterior distributions (orange; 50 samples) for phase-folded TESS photometry of TOI-1272b. Expected transit shape for a circular orbit shown in blue, modeled using median posterior distribution values for all other model parameters. Details regarding fitting procedure are discussed in \S\ref{sec:transit-model}. Bottom: Residuals to our maximum \emph{a posteriori} model.}
\label{fig:transit-model}
\end{figure}

We accessed the Pre-search Data Conditioning Simple Aperture Photometry (PDC-SAP; \citealt{Stumpe12}; \citealt{Stumpe14}; \citealt{Smith12}) through the Mikulski Archive for Space Telescopes (MAST), stitching together the light curves from individual TESS sectors into a single time-series using \texttt{Lightkurve} (\citealt{lightkurve18}). We performed outlier rejection, normalization, and de-trending of the full light curve following the procedures outlined in \cite{MacDougall21}. We then searched for transits using a box least squares (BLS; \citealt{Kovacs02}) transit search to recover the same planetary signal detected by SPOC with SNR = 23.3. We subtracted the known transits and applied the BLS search again but identified no additional periodic transit events.

To confirm that the observed transit events were on target and not the result of a background source, we referenced additional ground-based time-series photometry taken for TOI-1272. Independent observations were collected with MuSCAT2 (\citealt{Narita19}) at a pixel scale of 0\farcs{}44 in $g$, $r$, $i$, and $z_s$ filters on UT 2020 February 28 and again $\sim$1 year later with MuSCAT (\citealt{Narita15}) at a pixel scale of 0\farcs{}36 in $g$, $r$, and $z_s$ filters on UT 2021 May 8. These detections confirmed that the expected transit was on target and presented no evidence of nearby eclipsing binaries. This target has one neighbor listed in \emph{Gaia} Data Release 2 (DR2) within 30\arcsec{}. At a separation of 8\farcs{}45 and $\Delta$G = 5.93, the neighbor contributes $< 1\%$ dilution to the light curve, which was already corrected for in the photometric data products that we used.



\begin{figure}[ht]
\centering
\includegraphics[width=0.46\textwidth]{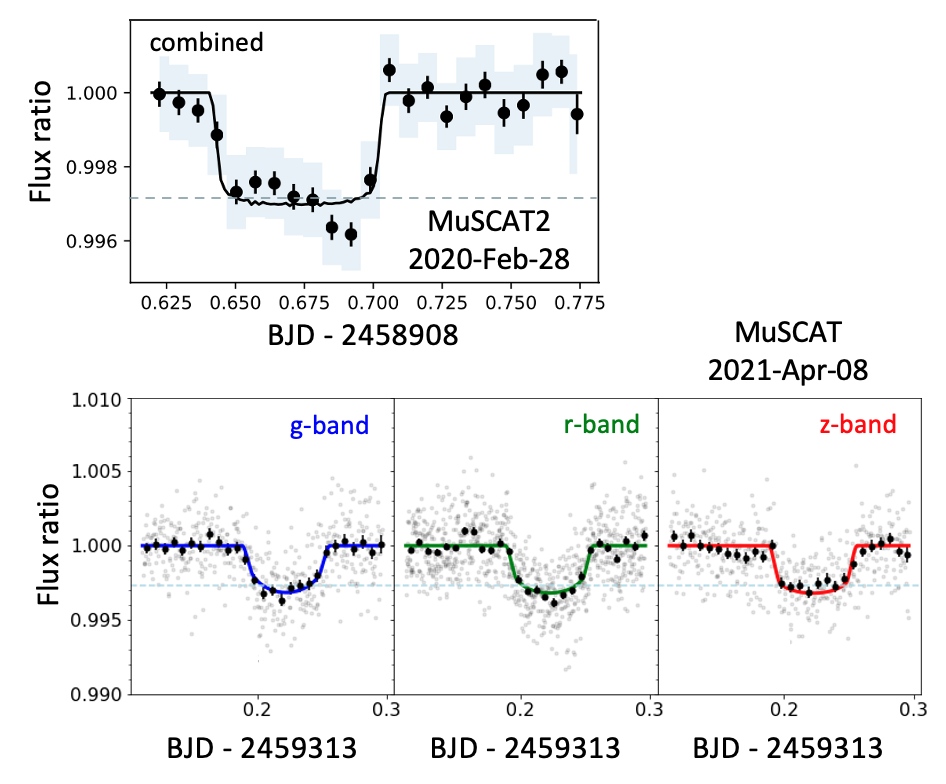}
\caption{Transit photometry for two independent single transits measured by the MuSCAT2 and MuSCAT instruments, plotted with 10 minute binning and a photometric fits by the corresponding instrument pipeline. Horizontal dashed lines indicate the expected transit depth. Top: Combined photometry from MuSCAT2 $g$, $r$, $i$, and $z_s$ bands, observed on UT 2020 February 28. Bottom: Photometry from MuSCAT $g$, $r$, and $z$ bands, observed on UT 2021 April 8.}
\label{fig:muscat}
\end{figure}

\subsection{Photometric Transit Model}
\label{sec:transit-model}


We characterized the planetary transit signal using a photometric light curve model to determine if TOI-1272b was a candidate for high-eccentricity. We made this determination by comparing the planet's observed transit duration ($T$; mid-ingress to mid-egress) to the expected duration for a circular orbit $T_{\rm circ}$. The ratio of these two values can be used to assess the orbital geometry of a transiting planet through the geometric relation for transit duration given by \citealt{Winn10}:

\begin{equation}
\label{eqn:duration}
T = \left(\frac{R_* P}{\pi a}\sqrt{1-b^2}\right)\frac{\sqrt{1-e^2}}{1+e\sin{\omega}},
\end{equation}

Given the known period $P = 3.316$ days, fixed $b = 0$, and the stellar characterization from \S\ref{sec:stellar-properties}, TOI-1272b would have a transit duration of $T_{\rm circ} = 0.094 \pm 0.004$ days if it were on a circular orbit. The observed transit, however, had a duration that was nearly 40\% shorter than this at $T_{\rm obs} \approx 0.06$ days. The short transit duration suggests either a high eccentricity orbit transiting near periastron or an orbit with a high impact parameter, motivating our follow-up analysis to constrain the true eccentricity.

\begin{figure}[ht]
\centering
\includegraphics[width=0.47\textwidth]{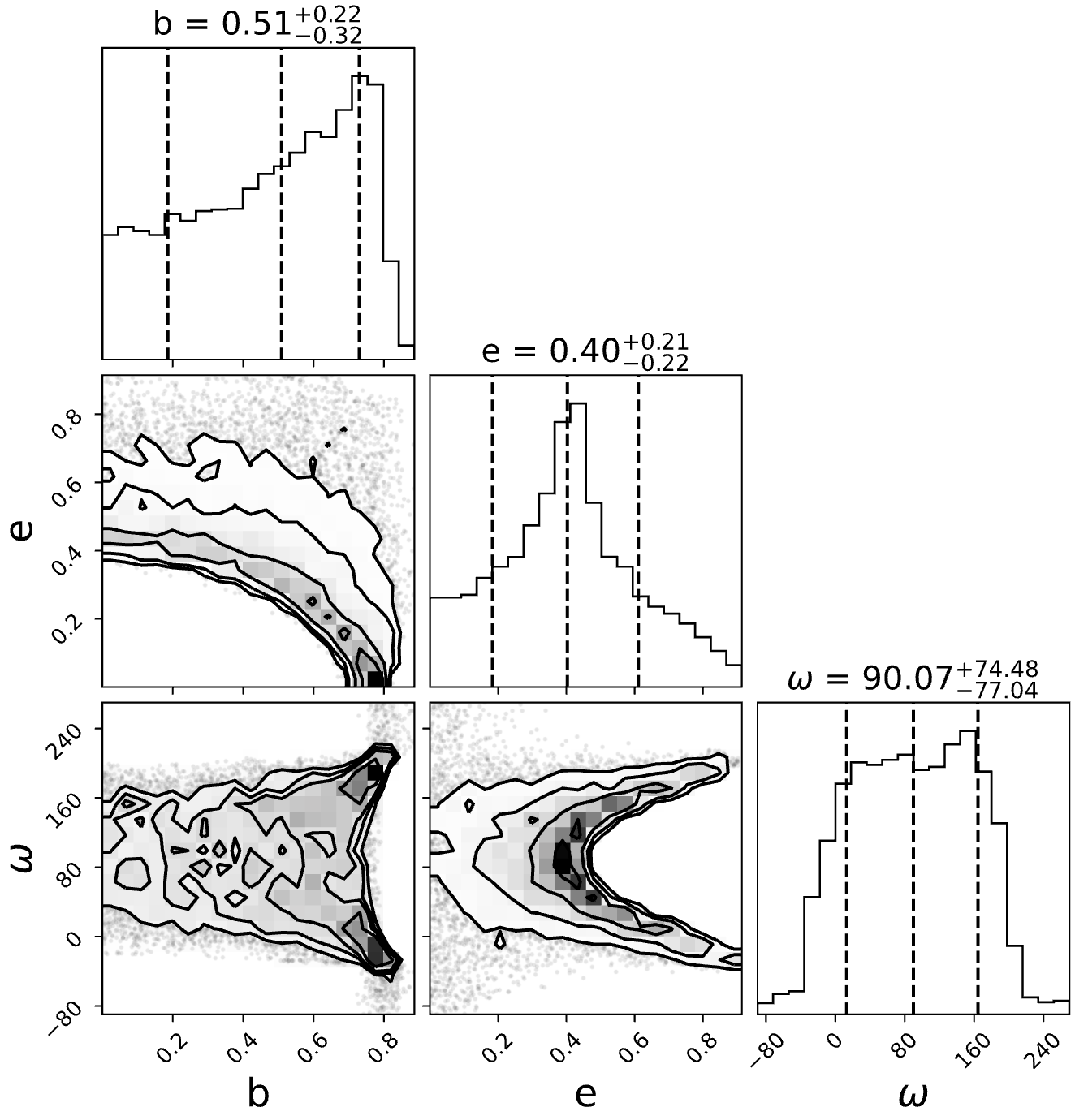}
\caption{Corner plot of \texttt{exoplanet} model posteriors for TOI-1272b, showing the effects of $e$-$\omega$-$b$ degeneracy on our transit fit. Best-fit value for eccentricity given by $e = 0.40^{+0.20}_{-0.22}$. Argument of periastron $\omega$ remained loosely constrained about 90$^{\circ}$, suggesting a transit near periastron.}
\label{fig:e_w-posterior}
\end{figure}

To characterize the transit properties of TOI-1272b more precisely, we fit the available TESS photometry with the \texttt{exoplanet} package (\citealt{Foreman-Mackey2021}). The \texttt{exoplanet} package uses a Hamiltonian Monte Carlo algorithm that is a generalization of the No U-Turn Sampling method (\citealt{Hoffman11}; \citealt{Betancourt16}). We used this model to generate samples from the posterior probability density for the parameters \{$P$, $t_0$, $R_p/R_*$, $b$, $\rho_{*}$, $\sqrt{e}\sin{\omega}$, $\sqrt{e}\cos{\omega}$, $\mu$, $u$, $v$\}, conditioned on the observed TESS light curve. Here, $\mu$ is the mean out-of-transit stellar flux and \{$u$, $v$\} are quadratic limb darkening parameters. The model used here follows that of \citealt{MacDougall21}. 

We applied weakly informative priors to each of the 10 model parameters, similar to those used by \cite{Sandford17}. In particular, the prior used on our parameterization of eccentricity and argument of periastron \{$\sqrt{e}\sin{\omega}$, $\sqrt{e}\cos{\omega}$\} was uniform on both parameters, not accounting for transit probability or other astrophysically motivated considerations. Also, our prior on $\rho_{*}$ was based on the stellar characterization discussed in \S\ref{sec:stellar-properties}. We fit the photometry of TOI-1272 with this model using 6,000 tuning steps and 4,000 sampling steps over 4 parallel chains. Figure \ref{fig:transit-model} shows the final transit model sampled from the posteriors.

An independent fit to the MuSCAT2 transit photometry of TOI-1272b was performed and used to verify the results of our transit fit to the full TESS photometry (Figure \ref{fig:muscat}). The raw MuSCAT2 data was reduced by the MuSCAT2 pipeline (\citealt{Parviainen19}) which performed standard image calibration, aperture photometry, and modeled the instrumental systematics present in the data while simultaneously fitting a transit model to the light curve. We also applied our own transit model to the detrended MuSCAT2 photometry, achieving consistent posterior constraints on all transit parameters. The same process was repeated for independent transit photometry from MuSCAT, producing similar results.

\begin{deluxetable}{lcrcc}
\tablewidth{0.99\textwidth}
\tabletypesize{\footnotesize}
\tablecaption{ Radial Velocity Measurements\label{tab:rv-data}}
\tablehead{
  \colhead{Time} & 
  \colhead{RV} & 
  \colhead{RV Unc.} & 
  \colhead{$S_{\rm HK}$} \\
  \colhead{(BJD)} & 
  \colhead{(m s$^{-1}$)} & 
  \colhead{(m s$^{-1}$)} & 
  \colhead{} 
}
\startdata
2458885.021035&11.00&1.68&0.315\\
2458904.925436&-0.66&1.55&0.277\\
2458911.896259&4.05&1.88&0.290\\
2458999.786695&4.78&1.51&0.304\\
2459003.840909&20.53&1.42&0.309\\
2459006.852128&-1.18&1.60&0.300\\
2459010.867973&4.47&1.55&0.294\\
2459024.775924&-3.86&1.58&0.300\\
2459024.829972&-3.96&1.36&0.295\\
2459024.883105&-6.08&1.70&0.302\\
\enddata
\tablecomments{Only the first 10 Keck/HIRES RVs are displayed in this table. A complete list has been made available online. $S_{\rm HK}$ values were measured using procedures from \cite{Isaacson10} with standard uncertainties of 0.001.}
\end{deluxetable}

\subsection{Eccentricity Constraints from Photometry}
\label{sec:photo-ecc}

The photometrically-constrained eccentricity posterior distribution that we measured for TOI-1272b from \{$\sqrt{e}\sin{\omega}$, $\sqrt{e}\cos{\omega}$\} was consistent with our high-eccentricity hypothesis, yielding a $1\sigma$ range of $e =$ 0.18--0.60 and an $\omega$ suggestive of a transit near periastron. The individual posterior distributions of $e$ and $\omega$ are shown in Figure \ref{fig:e_w-posterior}, along with their joint 2D posterior. 

We do note, however, that our impact parameter distribution remains loosely constrained, with a $1\sigma$ range of $b =$ 0.19--0.73, peaking in density towards the upper end of this range (Figure \ref{fig:e_w-posterior}). The similarly loose constraints on both $e$ and $\omega$ implied that our photometric characterization of the orbital geometry was complicated by $e$-$\omega$-$b$ degeneracy, as can be seen in the 2D joint posterior distributions in Figure \ref{fig:e_w-posterior}. Nevertheless, the potential for a high eccentricity combined with the expected Neptune-like size of the planet and its short orbital period made TOI-1272b a prime target for follow-up radial velocity observations.


\begin{figure*}[ht]
\centering
\includegraphics[width=0.85\textwidth]{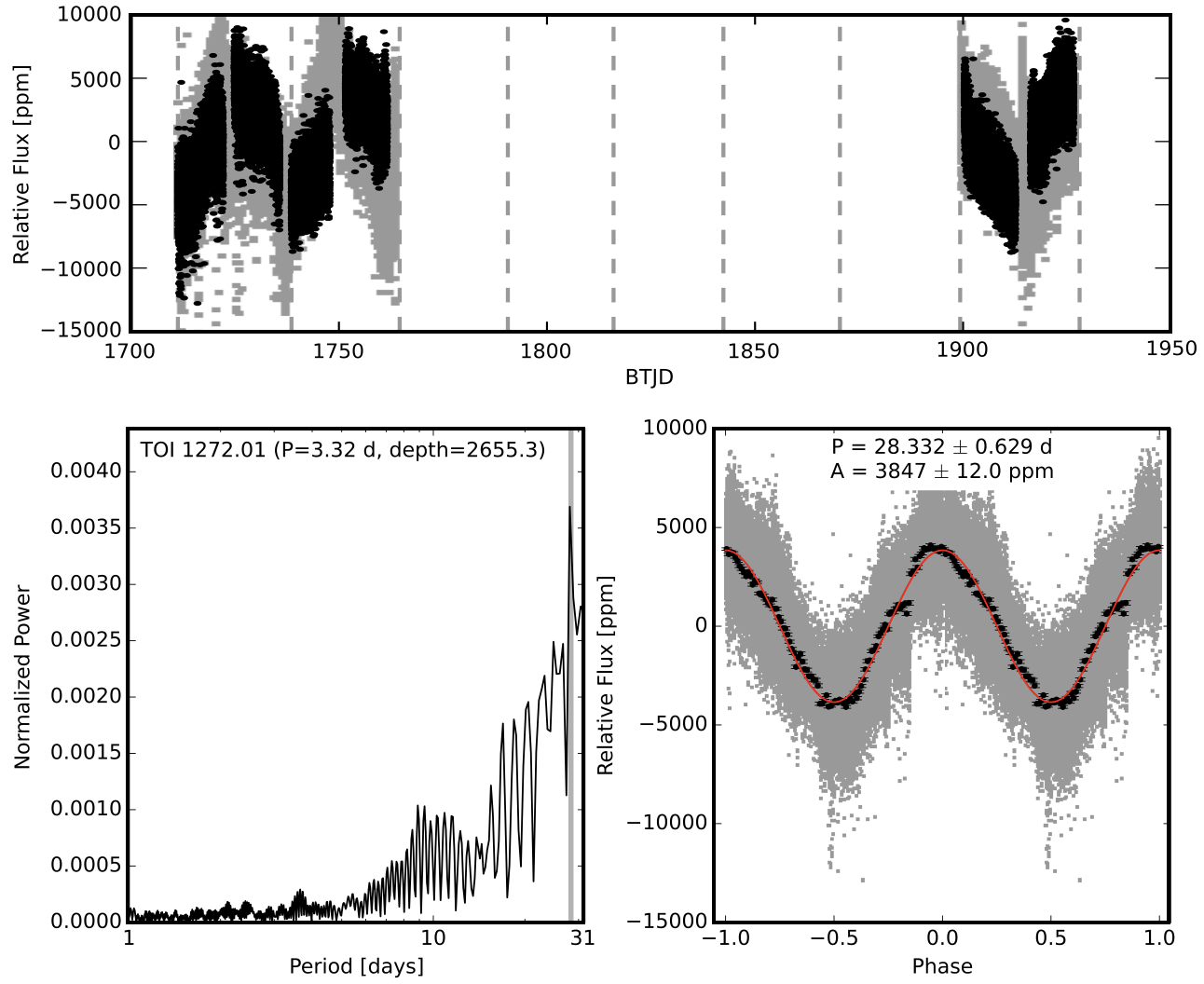} 
\caption{Top: TESS light curve of TOI-1272 from sectors 15, 16, and 22. The gray points show the original TESS SAP light curve and the black points show the \texttt{TESS-SIP} corrected light curve that is used to extract the variability period. The vertical dashed lines mark the gaps between TESS sectors. Bottom Left: Lomb-Scargle periodogram of the \texttt{TESS-SIP} corrected light curve. The orbital period and transit depth from the TOI catalog are listed at the top of the panel. Bottom Right: Phase-folded light curve based on the most significant period detected from the periodogram. The gray points show the \texttt{TESS-SIP} corrected photometry, the black points show the binned data, and the red curve is a sinusoidal fit to the phase-folded light curve. The period, amplitude, and their associated uncertainties from the best-fit sinusoidal function are listed at the top of the panel.
}
\label{fig:rotation}
\end{figure*}

\section{Spectroscopic Follow-up}
\label{sec:rvs}

We obtained a high-SNR template spectrum of TOI-1272 with the HIRES instrument at the Keck Observatory (\citealt{Vogt94}) on UT 2020 June 11 with 282 SNR pixel$^{-1}$ at 5000 \AA{}. We also collected 62 spectra of TOI-1272 between UT 2020 Feb 5 and UT 2021 November 27 (Table \ref{tab:rv-data}). On average, the observations had a spectral resolution of $R$ = 50,000, using a median exposure time of 900 s at 5500 \AA{}. Along with the RVs, we also measured the stellar activity S-index $S_{\rm HK}$ for all 62 Keck/HIRES observations using the observed strengths of the Ca II H and K lines in our spectra, following the methods of \cite{Isaacson10}.

For the RV observations, a heated cell of gaseous iodine was included along the light path just behind the entrance slit of the spectrometer, imprinting a dense forest of molecular absorption lines onto the observed stellar spectrum (\citealt{Marcy92}). These lines served as a wavelength reference for measuring the relative Doppler shift of each spectrum and tracking variations in the instrument profile using the standard forward-modeling procedures of the California Planet Search (\citealt{Howard10}).

\section{Stellar Characterization}
\label{sec:stellar}

\subsection{Bulk Properties}
\label{sec:stellar-properties}

Following the procedures outlined by \cite{MacDougall21}, we characterized the bulk properties of TOI-1272 by first inferring $T_{\rm eff}$ and [Fe/H] from our Keck/HIRES template spectrum using \texttt{SpecMatch-Synth} (\citealt{Petigura17b}). We then modeled the stellar mass, radius, surface gravity, density, and age via stellar isochrone fitting with \texttt{isoclassify} (\citealt{Berger20a}; \citealt{Huber17}). We report these values and their associated uncertainties in Table \ref{tab:system-properties}, accounting for small corrections due to model grid uncertainties discussed by \cite{Tayar20}. We note that the properties derived with \texttt{isoclassify} rely on \emph{2MASS} K-band magnitude and \emph{Gaia} parallax, also reported in Table \ref{tab:system-properties}. The properties that we measured were consistent with those reported to ExoFOP-TESS from two spectra obtained with the TRES instrument at the Whipple Observatory, analyzed using the Stellar Parameter Classification (SPC) tool (\citealt{Buchhave12}; \citealt{Buchhave14}).

\subsection{Variability and Rotation}
\label{sec:stellar-rotation}

To properly detrend our RV data and interpret any planetary signals, we first needed to characterize the intrinsic stellar variability for TOI-1272. We measured stellar variability from the TESS 2-min cadence SAP photometry where TOI-1272 was observed in 3 sectors, one of which partially overlapped with our RV observation baseline. Upon removing data that were flagged as being poor quality, $\geq$5$\sigma$ outliers, or part of the TOI-1272b transit events, we measured the stellar variability period from the trimmed SAP light curve using the \texttt{TESS-SIP} algorithm (\citealt{Hedges20}). This systematics-insensitive Lomb-Scargle periodogram (\citealt{Angus16}) yielded a clear variability signal at $28.3 \pm 0.6$ days, likely associated with stellar rotation. The corrected \texttt{TESS-SIP} light curve, Lomb-Scargle periodogram, and phase-folded light curve with a sinusoidal fit are shown in Figure \ref{fig:rotation}.

The stellar variability observed in the TESS photometry of TOI-1272 also allowed us to derive the expected stellar activity-driven variability in our RV measurements using the $FF^\prime$ method (\citealt{Aigrain12}). This method uses the light curve flux ($F$), its derivative ($F^\prime$), an estimate of relative spot coverage ($f \sim$ 0.005 in this work), and a simple spot model to simulate activity-induced RV variability. We estimated that this stellar activity would produce an RV variability signal with semi-amplitude $K \approx 5.0$ m s$^{-1}$, assuming a sinusoidal signal. Given the partial overlap of our RV baseline with that of the TESS photometry, we used this RV variability estimate as the foundation for our consideration of activity-driven RV signals in \S\ref{sec:other_signals}. By doing so, we implicitly assumed that the variability signal in the region of overlap could be extrapolated out to the entire data set. Based on the measured value of log($g$) and the observed values of activity metrics $S_{\rm HK}$ and log$R'_{\rm HK}$, moderate stellar activity-driven RV jitter was also expected for TOI-1272 based on the classifications presented in \citealt{Luhn20}, $\sigma_{\rm jit} \gtrsim 2.5$ m s$^{-1}$, consistent with our photometry-only estimate.

\begin{figure}[ht]
\centering
\includegraphics[width=0.45\textwidth]{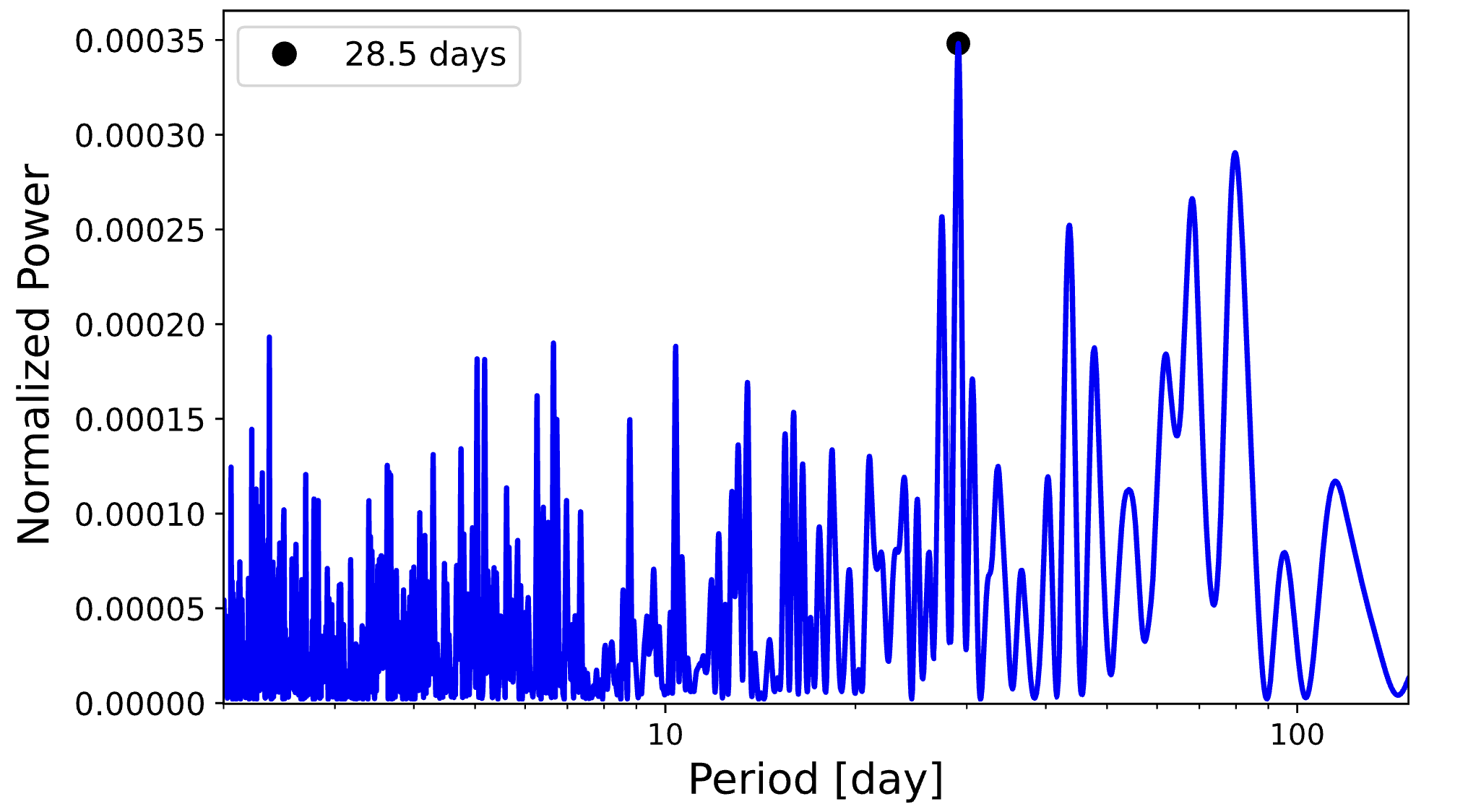}
\caption{Lomb-Scargle periodogram search of TOI-1272 $S_{\rm HK}$ data using, identifying a significant 28.5 day variability signal consistent with the suspected stellar rotation period.}
\label{fig:svals}
\end{figure}

As a final consideration of stellar variability, we searched for periodic, activity-driven signals in the $S_{\rm HK}$ data series for TOI-1272 using a Lomb-Scargle periodogram. We identified a significant 28.5 day signal, consistent with our stellar variability measurement from \texttt{TESS-SIP} (Figure \ref{fig:svals}). We also detected additional sub-significant $S_{\rm HK}$ variability signals that did not correspond to any known sources. We consider the impact that activity may have on our RV measurements when constructing our RV-only model in \S\ref{sec:keplerian-model}.

\subsection{Age}
\label{sec:stellar-age}
Given the short orbital period of TOI-1272b and the possibility of a high-eccentricity orbit, the age constraints for this system were valuable for interpreting the tidal circularization timescale of the transiting planet. Our isochrone fit using \texttt{isoclassify} yielded a poorly constrained age estimate of $\sim$1--7 Gyr. This was consistent with a first-order analytical estimate of the age of TOI-1272, 3.1 Gyr, based on $G_{\rm BP} - G_{\rm RP}$ color and stellar rotation period via gyrochronology (\citealt{Angus19b}). 

We took this analysis a step further by using \texttt{stardate} (\citealt{Angus19}) to combine stellar isochrone modeling with gyrochronology to precisely measure the stellar age. Running the \texttt{stardate} MCMC sampler for $10^5$ draws, we measured an age of $3.65\substack{+4.17 \\ -0.98}$ Gyr. While this age range remained broad and consistent with our \texttt{isoclassify} measurement, the increased median value and reduced lower uncertainty from \texttt{stardate} provided us with better constraints on the lower bound of the age of this system.

\section{Keplerian Modeling}
\label{sec:keplerian-model}

\subsection{RV Detection of Planets b and c }
\label{sec:detection}

\begin{figure}[ht]
\centering
\includegraphics[width=0.4\textwidth]{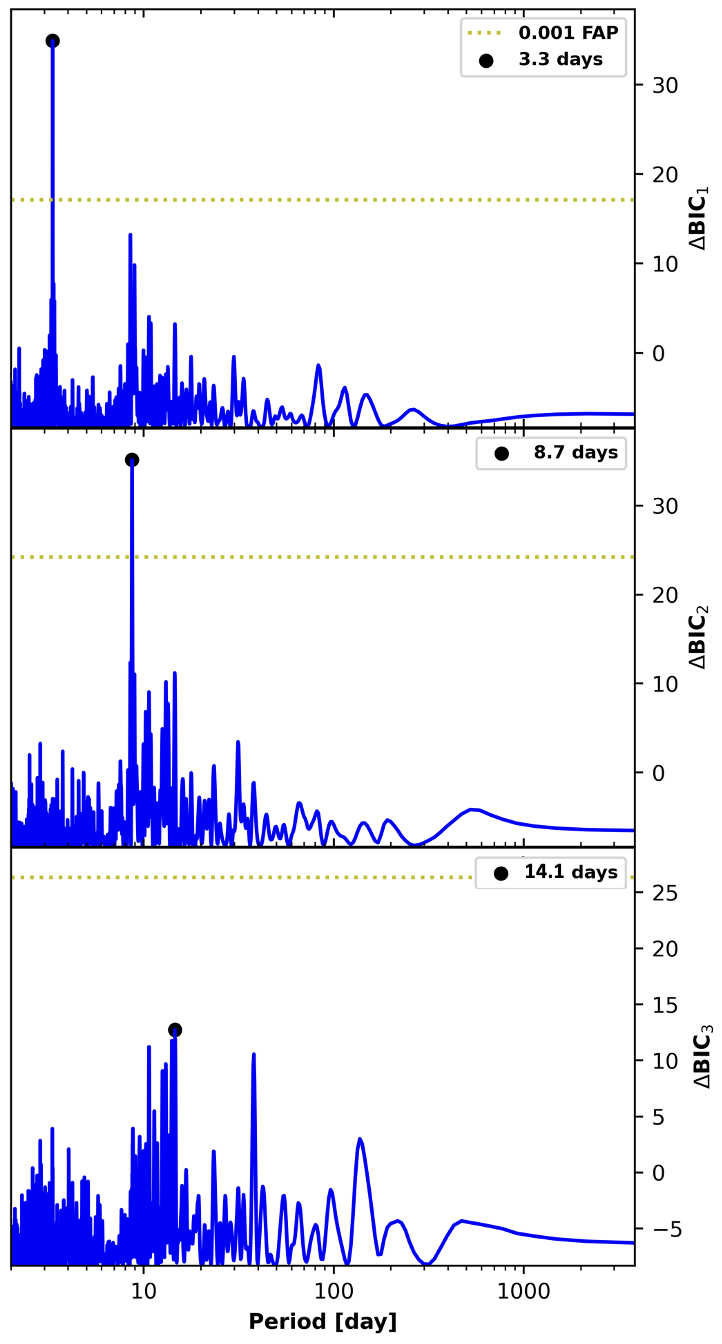}
\caption{Iterative Keplerian periodogram search of TOI-1272 RV data using \texttt{RVSearch}. We confirmed the 3.3-day transiting planet (panel a) and identified a significant 8.7-day period with no corresponding transits (panel b). $\Delta$BIC was used to discriminate between models with additional Keplerians over a grid of periods (Bayesian Information Criterion; \citealt{Schwarz78}), corresponding to a significance threshold of FAP = 0.001 at the yellow dashed horizontal line. Panel c shows a sub-significant signal at a 14.1 day period, likely corresponding to $P_{\rm rot}/2$.}
\label{fig:rv-search}
\end{figure}

We searched for periodic signals in our RV data using the \texttt{RVSearch} pipeline (\citealt{Rosenthal21}). We set Gaussian priors on the period $P_b$ and time of conjunction $T_{\rm c,b}$ of the 3.3-day planetary signal known from photometry. We then used \texttt{RVSearch} to iteratively search the RV data for additional Keplerian signals across the period range from 2 to 4000 days. This search yielded an eccentric Keplerian fit with $K \approx 12.6$ m s$^{-1}$ at the known period and an outer 8.7-day Keplerian fit with $K \approx 9.4$ m s$^{-1}$ (Figure \ref{fig:rv-search}). Both signals surpassed our significance threshold, with false-alarm probabilities (FAP) measured by \texttt{RVSearch} FAP $\approx$ $10^{-4}$ and $10^{-5}$, respectively. 

We corroborated the significance of the 8.7-day signal by performing an independent search of the RV data set using an $l_1$ periodogram (\citealt{Hara17}), which minimizes the aliasing seen in a general Lomb-Scargle periodogram by evaluating all frequencies simultaneously rather than iteratively. We implemented our $l_1$ periodogram with jitter $\sigma = 5.0$ m s$^{-1}$, correlation time $\tau = 0$, and maximum frequency 1.5 cycles d$^{-1}$ across the period range from 1.1 to 1000 days. Both the 3.3 and 8.7 day signals were clearly detected by from this $l_1$ periodogram search, with consistent FAP values of $\sim$10$^{-4}$ and $\sim$10$^{-5}$, respectively. Given the significance of the 8.7-day period and the lack of a corresponding signal in either the $S_{\rm HK}$ activity data or photometric time series (see \S\ref{sec:stellar-rotation}), we concluded that this Keplerian signal was of planetary origin. A close inspection of the phase-folded and detrended TESS photometry at the RV-constrained period and time of conjunction for the outer RV signal showed no evidence for a corresponding transit event.

\subsection{Additional RV Signals}
\label{sec:other_signals}

While we did not identify any additional signals in our RV data that met our significance criteria, we did detect a sub-significant Keplerian signal at a 14.1 day period using both \texttt{RVSearch} and an $l_1$ periodogram search. This signal persisted throughout our entire observing baseline and was detectable in the residuals to a preliminary two-planet RV fit with Keplerian modeling code \texttt{RadVel} (\citealt{Fulton18a}). We concluded that this signal was the first harmonic ($P_{\rm rot}/2$) of the 28.3 day stellar rotation as measured from the $S_{\rm HK}$ time series and TESS photometry. The $P_{\rm rot}/2$ harmonic of a star's rotation period is known to induce strong periodic activity signatures such as this in RV time-series data (\citealt{Boisse11}).  

Preliminary RV modeling revealed no other RV signals and insignificant detections of a trend and curvature in our RV time-series, providing no evidence of further companions. We also found an estimated RV jitter of $\sigma \approx$ 5.5 m s$^{-1}$. This jitter measurement was consistent with both our $FF^\prime$ estimate of RV variability and the RV semi-amplitude of the marginal 14.1 day signal ($K = 4 \pm 1$ m s$^{-1}$). Given the low significance of this additional signal and its sub-jitter amplitude, we chose to only consider the two planetary signals in our final RV models. We therefore interpreted the spectroscopic data for TOI-1272 to reveal 2 planetary signals (3.3 days and 8.7 days), with a sub-significant activity signal driven by stellar rotation ($P_{\rm rot}/2 \approx$ 14.1 days).

\begin{figure}[ht]
\centering
\includegraphics[width=0.47\textwidth]{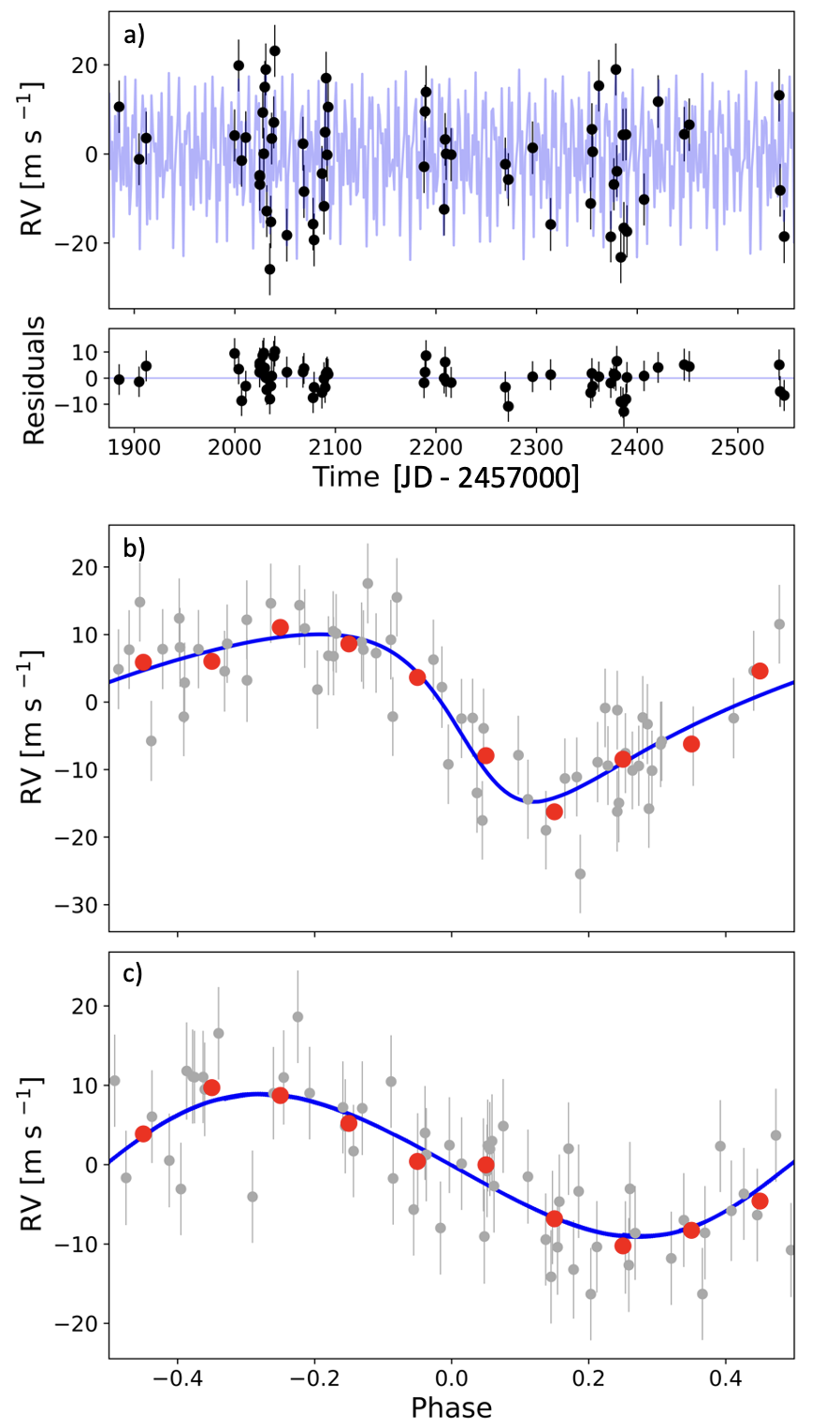}
\caption{(a) Best-fit radial velocity model (blue) for Keck/HIRES RV measurements (black) using \texttt{Radvel} (\citealt{Fulton18a}) in a joint RV-photometry model via \texttt{juliet} (\citealt{juliet}), with corresponding residuals shown below. (b)-(c) Phase-folded views of best-fit RV model for TOI-1272 b and c, with binned points shown in red.}
\label{fig:radvel-model}
\end{figure}

\subsection{RV-only Model}
\label{sec:rv-fit}

We performed a two-planet fit to the RV time series for TOI-1272 using \texttt{RadVel}, a Python package used to characterize planets from Keplerian RV signals by applying maximum \emph{a posteriori} model fitting and parameter estimation via MCMC (\citealt{Fulton18a}). Our model consisted of two planetary Keplerian signals with periods 3.3 and 8.7 days. We modeled the data by fitting the following free parameters for both planets: $P$, $T_c$, $K$, $\sqrt{e} \cos{\omega}$, and $\sqrt{e} \sin{\omega}$. Our model also included RV offset $\gamma$ and RV jitter term $\sigma$ to account for astrophysical white noise and instrumental uncertainty. The best-fit RV-only \texttt{RadVel} model confirmed the existence of two eccentric sub-Jovian mass planets orbiting TOI-1272, and we used these results to inform the priors for a joint RV-photometry model.

\subsection{RV-Photometry Joint Model}
\label{sec:joint-fit}

We obtained the most precise planet parameters for the TOI-1272 system by performing global RV-photometry modeling using \texttt{juliet} (\citealt{juliet}), a robust tool for modelling both transiting and non-transiting exoplanets. We used \texttt{juliet} to jointly fit the radial velocities through \texttt{RadVel} and the transit photometry through \texttt{batman} (\citealt{batman}), with proper handling of limb-darkening coefficients (\citealt{Kipping13}). Estimation and comparison of Bayesian evidences and posteriors was performed directly by the dynamic nested sampling package \texttt{dynesty} (\citealt{dynesty}), one of several such tools offered through the \texttt{juliet} interface. Unlike the Monte Carlo algorithm used in our initial transit-only analysis, nested sampling algorithms break up complex posterior distributions into simpler nested slices, sampling from each slice individually then recombining the weighted results to reconstruct the complete posterior. This method becomes more efficient in the higher-dimensional posterior spaces of joint models. 

We directly fit for each transit and Keplerian property with priors informed from our previous photometry-only and RV-only models. For the final global model, we fit for photometry-only properties \{$R_{\rm p}/R_*$, $b$, $\rho_{*}$, $\mu$, $u$, $v$\}, joint properties \{$P_{\rm b}$, $t_{\rm 0,b}$, $\sqrt{e_{\rm b}}\sin{\omega_{\rm b}}$, $\sqrt{e_{\rm b}}\cos{\omega_{\rm b}}$\}, and RV-only properties \{$P_c$, $t_{\rm 0,c}$, $\sqrt{e_{\rm c}}\sin{\omega_{\rm c}}$, $\sqrt{e_{\rm c}}\cos{\omega_{\rm c}}$, $K_{\rm b}$, $K_{\rm c}$, $\gamma$, $\sigma$\}. Our final measurements are included in Table \ref{tab:system-properties} and the corresponding maximum \emph{a posteriori} RV model is shown in Figure \ref{fig:radvel-model}.

In summary, we measured mass constraints for TOI-1272 b and c at significance levels $\sim$11$\sigma$ and $\sim$9$\sigma$, respectively, reflecting the strengths of the two periodogram signals discussed in \S\ref{sec:detection}. We also measured a high eccentricity of $e_b = 0.34 \pm 0.06$ for TOI-1272b, consistent within $1\sigma$ of our photometry-only eccentricity constraint from \S\ref{sec:photo-ecc}. The eccentricity of the outer planet was loosely constrained to $e_c = 0.12\substack{+0.1 \\ -0.08}$. We note, however, that a model fit with $e_c = 0$ performed nearly identically to the eccentric model, suggesting that the eccentricity of TOI-1272c is only marginally significant. We discuss these constraints on eccentricity further in \S\ref{sec:stability}. Our global model also served to minimize degeneracies between $e$-$\omega$-$b$ and allowed us to obtain more precise $b$ and $R_{\rm p,b}$ values than with our photometry-only model. Our loose posterior constraint on impact parameter from Figure \ref{fig:e_w-posterior} was improved to $b = 0.45^{+0.15}_{-0.21}$, subsequently yielding our final radius measurement of $R_{\rm p,b} = 4.14$ $\pm$ 0.21 $R_\earth$.

\begin{deluxetable}{lrc}
\tablewidth{0.99\textwidth}
\tabletypesize{\footnotesize}
\tablecaption{TOI-1272 System Properties
\label{tab:system-properties}}
\tablehead
{
  \multicolumn{1}{l}{Parameter}&
  \multicolumn{1}{r}{Value}&
  \multicolumn{1}{c}{Notes}
}

\startdata
$Stellar$   &   &   \\
RA ($^{\circ}$)	& 199.1966	& A \\
Dec ($^{\circ}$)	& 49.86104	& A \\
$\pi$ (mas)	& $7.24 \pm 0.021$	& A \\
$m_G$	& $9.6844 \pm 0.0004$	& A \\
$m_K$	& $9.70 \pm 0.02$	& B \\
$T_{\rm eff}$ (K)	& $4985 \pm 121$	& C \\
$[$Fe/H$]$ (dex)	& $0.17 \pm 0.06$	& C \\
log($g$)	& $4.55 \pm 0.10$	& C \\
$M_*$ ($M_\odot$)	& $0.851 \pm 0.049$	 & C \\
$R_*$ ($R_\odot$)	&	 $0.788 \pm 0.033$	& C \\
$\rho_*$ (g cm$^{-3}$)	& $2.453 \pm 0.343$	& C \\
age (Gyr)	& $3.65\substack{+4.17 \\ -0.98}$ & D \\
$P_{\rm *,rot}$ (days)	& $28.3 \pm 0.6$ & E \\
$S_{\rm HK}$	& $0.331$	& F \\
log$R'_{\rm HK}$	& $-4.705$	& F \\
$u$	& $0.39 \pm 0.05$	 & G \\
$v$	& $0.09 \pm 0.05$	 & G \\
$\mu$ (ppm)	& $-4.3 \pm 36.2$	& H \\
$\gamma$ (m s$^{-1}$)	& $0.7 \pm 0.7$	& H \\
$\sigma_{\rm jit}$ (m s$^{-1}$)	& $5.6 \pm 0.6$	& H \\
\\
$Planet$ $b$    &   &   \\
$P$ (days)	& $3.31599 \pm 0.00002$	& H \\
$T_c$ (BJD-2457000)	& $1713.0253 \pm 0.0006$	& H \\
$b$	& $0.45\substack{+0.15 \\ -0.21}$	 & H \\
$R_p$ ($R_\earth$)	& $4.14 \pm 0.21$	& H \\
$M_p$ ($M_\earth$)	& $24.6 \pm 2.3$	& H \\
$\rho_p$ (g cm$^{-3}$)	& $1.9 \pm 0.3$	& H \\
$K$ (m s$^{-1}$)	& $12.6 \pm 1.1$	& H \\
$a$ (AU)	&	 $0.0412 \pm 0.0008$	& H \\
$e$	& $0.338\substack{+0.056 \\ -0.062}$	& H \\
$\omega$ ($^{\circ}$)	& $123.6 \pm 11.5$	& H \\
$T_{\rm eq}$ (K)	& $961 \pm 32$	& I \\
\\
$Planet$ $c$    &   &   \\
$P$ (days)	& $8.689 \pm 0.008$	& H \\
$T_c$ (BJD-2457000)	& $1885.34 \pm 0.48$	& H \\
$M_p$ sin$i$ ($M_\earth$)	& $26.7 \pm 3.1$	& H \\
$K$ (m s$^{-1}$)	& $9.4 \pm 1.0$	& H \\
$a$ (AU)	&	 $0.0783 \pm 0.0014$	& H \\
$e$	& $\lesssim 0.35$	& J \\
$\omega$ ($^{\circ}$)	& $-80.8\substack{+97.4 \\ -57.3}$	& H \\
$T_{\rm eq}$ (K)	& $697 \pm 23$	& I \\
\enddata
\tablecomments{A: \emph{Gaia} DR2, epoch J2015.5 (\citealt{GaiaDR2}); B: \emph{2MASS} (\citealt{Skrutskie06}); C: Derived with \texttt{isoclassify}; D: Derived with \texttt{stardate} (\citealt{Angus19}); E: Derived with \texttt{TESS-SIP} (\citealt{Hedges20}); F: Measured from Keck/HIRES template; G: Derived with \texttt{LDTK} (\citealt{ldtk15}); H: Constrained from joint RV-photometry model with \texttt{juliet} (\citealt{juliet}; \citealt{batman}, \citealt{Fulton18a}, \citealt{dynesty}); I: Calculated from other parameters assuming albedo $\alpha = 0.3$; J: Dynamically constrained with \texttt{rebound} (\citealt{Rein12}).}
\end{deluxetable}

\section{System Dynamics}
\label{sec:dynamics}

\subsection{Eccentricity Constraints from Stability Requirements}
\label{sec:stability}

Despite the compact architecture of the TOI-1272 system, both planets had moderate RV-constrained eccentricities that were inconsistent with zero to $\sim$5$\sigma$ and $\sim$1$\sigma$ significance, respectively. Such excited dynamics put TOI-1272 b and c at risk of dynamical instability if orbit crossing were to occur: 

\begin{equation}
\label{eqn:amd}
\frac{a_c\left(1-e_c\right)}{a_b\left(1+e_b\right)} > 1.
\end{equation}

Given our RV-constrained measurements of eccentricity and orbital separation in this system, we found the left-hand side of the above equation to be $1.22 \pm 0.15$, or <2$\sigma$ from the orbit crossing threshold. We note, however, that our confidence in long-term orbital stability based on this value is highly sensitive to our eccentricity uncertainties. Assuming a fixed true value of $e_c = 0.05$, this system would be firmly out of reach of geometric orbit crossing given its current configuration.

With the ambiguity in our orbit-crossing stability result, we also calculated the dynamical stability of the TOI-1272 system according to the stricter criterion from \cite{Petrovich15}:
\begin{equation}
\label{eqn:petrovich}
\frac{a_c\left(1-e_c\right)}{a_b\left(1+e_b\right)} - 2.4\left(\rm max\left(\mu_b,\mu_c\right)\right)^{1/3}\left(\frac{a_c}{a_b}\right)^{1/2} > 1.15,
\end{equation}
where $\mu$ is $M_p$/$M_*$, drawn from our joint model results. This threshold marks an estimated empirical boundary in two-planet system stability, determined by applying a Support Vector Machine algorithm to a large number of numerical integrations. Planet-planet interactions resulting in ejecting a planet into the star or out of the system were considered by \cite{Petrovich15} in developing this criterion. 

Systems that satisfy the condition in Eq. \ref{eqn:petrovich} are expected to maintain dynamical stability for integrations out to at least $10^8$ orbits of the inner planet. When computed for this system, we measured the left-hand side of Eq. \ref{eqn:petrovich} to be $1.1 \pm 0.15$, or <1$\sigma$ below the stated stability threshold of 1.15. Similar to the orbit crossing criterion, TOI-1272 straddles the stability boundary for the \cite{Petrovich15} empirical threshold. We again note that a fixed outer planet eccentricity of $e_c = 0.05$ would promote the long-term stability of the TOI-1272 system according this stability criterion.

\begin{figure}[ht]
\centering
\includegraphics[width=0.45\textwidth]{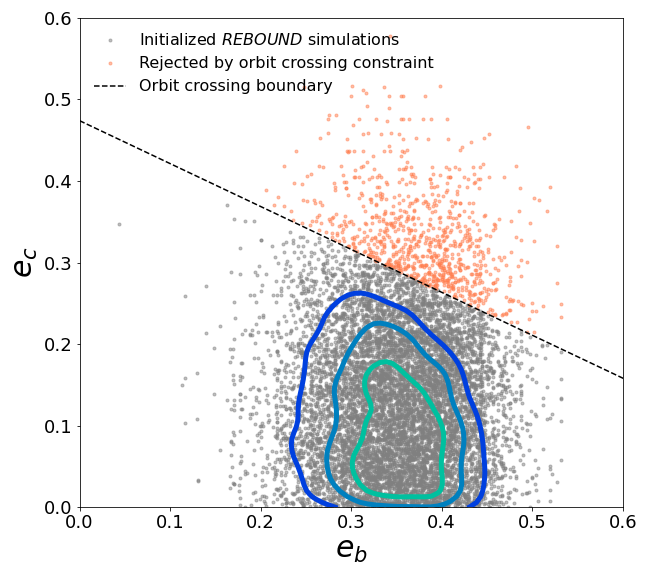}
\caption{Distribution of initialized $e_b$-$e_c$ values for all \texttt{rebound} simulations of the TOI-1272 system. Green-blue contours show regions with highest density of stable configurations (simulations lasting $10^6$ orbits of the inner planet).}
\label{fig:rebound}
\end{figure}

We followed up these inconclusive analytical predictions of long-term stability with a full N-body treatment of the stability of the TOI-1272 system. Drawing initial conditions from our RV-photometry model posteriors, we ran $10^4$ N-body simulations with \texttt{rebound} (\citealt{Rein12}) for $\sim$10$^6$ orbits of the inner planet. We restricted the initialized eccentricities of our simulations to avoid starting on crossing orbits, and we considered a simulation to be "unstable" after an orbit-crossing event or close dynamical encounter. Overall, $\sim$81$\%$ of simulations remained "stable" for the entirety of our integration time, suggesting that the eccentricities and masses measured from our RV model were largely consistent with a stable architecture on moderate time-scales (Figure \ref{fig:rebound}). Our \texttt{rebound} simulations also showed that stable configurations of this system exhibit Laplace-Lagrange oscillations in eccentricity with a secular timescale on the order of $\sim$10$^2$ years. 

While our $e_b$ posterior remained mostly unchanged by this N-body model, our stability constraints on $e_c$ allowed us to determine an upper bound of $e_c \lesssim 0.35$. Upon this redefinition of $e_c$, we inferred that the true eccentricity of TOI-1272c was likely in the lower tail of the acceptable range. Given that a \texttt{RadVel} model with $e_c = 0$ performed nearly equivalently to the non-zero eccentricity model (\S\ref{sec:joint-fit}), this interpretation is consistent with our RV-only analysis.

\subsection{No Evidence for TTVs}
\label{sec:ttvs}

Along with our N-body integration, we also used \texttt{rebound} to model the transit-timing variations (TTVs) expected to be observed in this system over a similar baseline as our TESS photometry ($\sim$215 days). We estimated a TTV O-C RMS of $0.3 \pm 0.3$ minutes, below our threshold of sensitivity for individual transits. We verified this empirically by modeling the TESS photometry with \texttt{exoplanet}, similar to \S\ref{sec:photo-ecc} but this time including TTVs as an additional model parameter. From this fit, we measured a TTV O-C RMS of $2 \pm 2$ minutes, consistent with the estimate from \texttt{rebound}. 

With the additional photometric observations from MuSCAT, we extended our TTV search to a total photometric baseline of $\sim$600 days. A fit to the single transit measured by MuSCAT yielded a transit mid-point of BJD-2457000 = 2313.223, only $\sim$3 minutes off from predicted mid-point of $2313.221 \pm 0.003$ and within the $\sim$5 minute uncertainty of this prediction. The extended photometric baseline demonstrated again that the TTVs in this system are negligible, within $\sim$1$\sigma$ of showing no evidence for TTVs. 

The lack of TTVs was also consistent with the non-resonant orbital period ratio between TOI-1272 b and c: $P_c/P_b \approx 2.62$. This period ratio is outside of the resonant width of any strong resonances, lying most closely to the 3:1 second-order mean-motion resonance (MMR), with a $\sim$14$\%$ difference in ratio. However, we cannot rule out the possibility of past planet migration leading to resonance-crossing, which could have played a role in the planet-planet excitation of $e_b$ discussed briefly in \S\ref{sec:formation}.

\subsection{Strong Tidal Eccentricity Decay}
\label{sec:tides}

The age measurement of TOI-1272 from \S\ref{sec:stellar-age} is valuable when considering the potential tidal eccentricity decay of TOI-1272b. According to \cite{Millholland20}, which draws from \cite{Leconte10}, the timescale of orbital circularization due to tidal eccentricity damping for an eccentric orbit is given by

\begin{multline}
\label{eqn:tides}
\tau_e = \frac{4}{99}\left(\frac{Q'}{n}\right)\left(\frac{M_P}{M_*}\right)\left(\frac{a}{R_P}\right)^5 \\
\times\left(\Omega_e\left(e\right)\cos{\epsilon}\left(\frac{\omega_{\rm eq}}{n}\right)-\frac{18}{11}N_e\left(e\right)\right)^{-1}.
\end{multline}
Here, the mean motion is given by $n = \sqrt{G M_* / a^3}$ and the reduced tidal quality factor $Q'$ can be rewritten as $Q' = 3 Q / 2 k_2$, with specific dissipation function $Q$ and tidal Love number $k_2$ (\citealt{Murray99}; \citealt{Mardling04}). We defined $\omega_{\rm eq}$ as the spin rotation frequency of TOI-1272b at equilibrium, which we found to be $3 \pm 0.5$ day$^{-1}$ following the procedure outlined in \cite{Millholland20}. We assumed the obliquity $\epsilon$ to be $0^{\circ}$. We have also introduced functions of eccentricity $\Omega_e(e)$ and $N_e(e)$ given by

\begin{equation}
\label{eqn:Ne}
\Omega_e\left(e\right) = \frac{1+\frac{3}{2}e^2+\frac{1}{8}e^4}{\left(1-e^2\right)^5}
\end{equation}
\begin{equation}
\label{eqn:Omegae}
N_e\left(e\right) = \frac{1+\frac{15}{4}e^2+\frac{15}{8}e^4+\frac{5}{64}e^6}{\left(1-e^2\right)^{\frac{13}{2}}}.
\end{equation}

A typical Neptune-like planet is generally assumed to have a tidal quality factor of $Q' \approx 10^5$, but the true value is highly uncertain. Assuming this fixed value for $Q'$ and drawing the other parameters in Eq. \ref{eqn:tides} from our previous analysis, we estimated a circularization timescale of $\tau_e \approx 0.21 \pm 0.09$ Gyr. This nominal value of $\tau_e$ is >3$\sigma$ below our age measurement of $3.65\substack{+4.17 \\ -0.98}$ Gyr, suggesting that TOI-1272b has experienced significant eccentricity decay due to tides. This is not reflected in the anomalously high eccentricity that we measured, suggesting that another mechanism must be driving the excited state of this system. We note, however, that $Q'$ is highly uncertain and $\tau_e \propto Q'$, so a tidal quality factor of $2 \times 10^6$ would make $\tau_e$ consistent with the age of the system.

Continuing with the assumption of $Q' \approx 10^5$, we estimated the initial eccentricity that would have been needed for TOI-1272b to reach its currently observed $e_b$ after $3.65$ Gyr of tidal eccentricity decay. Assuming constant $Q'$ and $\tau_e$, we followed the procedures of \cite{Correia20} to derive the required post-formation eccentricity of $e_b \approx 0.8$. Without a significant restructuring of the TOI-1272 system architecture, however, such a high eccentricity would not have allowed for a stable companion at the orbital separation of TOI-1272c. Ruling out this "hot-start" scenario, we are left to consider whether the anomalously high eccentricity of the inner planet is due to an underestimated $Q'$ or excitation by some other dynamical mechanism. We discuss such formation and evolution scenarios further in \S\ref{sec:formation}.


\section{Context in the Hot Neptune Desert}
\label{sec:context}

\subsection{Bulk Density and Core-Envelope Fraction}
\label{sec:composition}

TOI-1272b is a Neptune-like planet for which we measured a mass of $24.6 \pm 2.3$ $M_{\rm \earth}$ and a radius of $4.1 \pm 0.2$ $R_{\rm \earth}$, yielding a density of $1.9 \pm 0.3$ g cm$^{-3}$. This system also contains a similar-mass outer companion ($M_p$ sin$i$ $= 26.7 \pm 3.1$ $M_\earth$) that is not transiting. Planets in this size and mass range have been reported frequently in the literature, only a few of which also fall within the Hot Neptune Desert (\citealt{Mazeh16}; \citealt{Owen18}). At moderate planet sizes ($\sim$2--6 $R_{\rm \earth}$; $M_p \lesssim 100$ $M_{\rm \earth}$) and low orbital separations ($P \lesssim 5$ days), a relative paucity of planets has been observed. 

\begin{figure*}[ht]
\centering
\includegraphics[width=0.95
\textwidth]{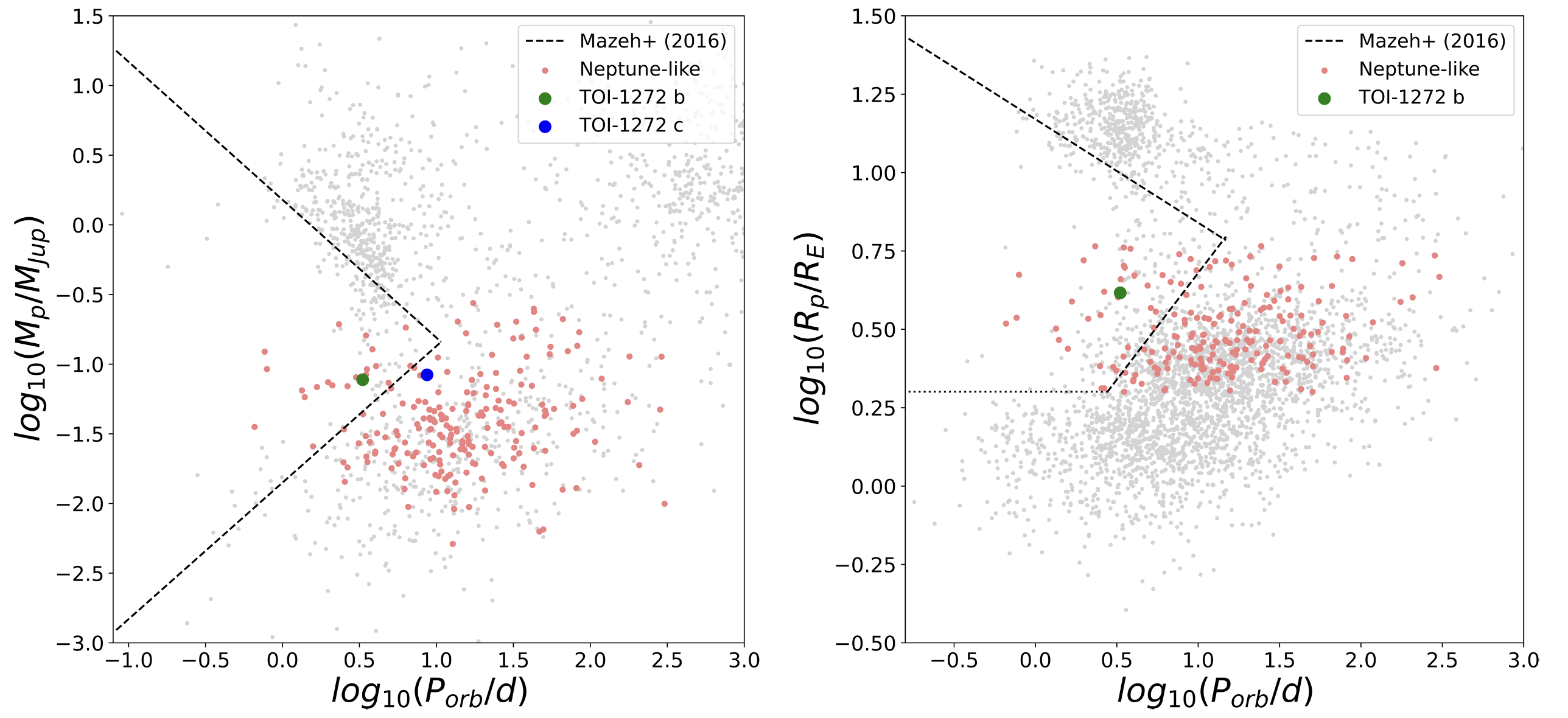}
\caption{Hot Neptune Desert in $R_p$-$P$ and $M_p$-$P$ parameter space, where Neptune-like is defined as $M_p < 100$ $M_{\rm \earth}$ and $R_p = 2.0-6.0$ $R_{\rm \earth}$. Data drawn from the NASA Exoplanet Archive (\citealt{Akeson13}; \citealt{ps}). TOI-1272 b and c depicted as green point (both panels) and blue point (left panel), respectively. Relations dictating the upper and lower boundaries of the Hot Neptune Desert from \cite{Mazeh16} shown as black dashed lines. Left: Mass-period distribution of all planets, with Neptune-like planets shown in red. Right: Radius-period distribution of all planets, with Neptune-like planets shown in red and $R_p = 2$ $R_{\rm \earth}$ limit shown as black dotted line.}
\label{fig:context}
\end{figure*}

The triangular regions of $R_p$-$P$ and $M_p$-$P$ parameter space shown in Figure \ref{fig:context} highlight this phenomenon, as defined by \cite{Mazeh16}. TOI-1272b can be seen here among the small subset of Neptunes that fall within this otherwise sparse parameter space. Some notable inhabitants of the Hot Neptune Desert include GJ 436b (\citealt{Lanotte14}) and HAT-P-11b (\citealt{Yee18}) along with more recent finds from TESS photometry including LP 714-47b (\citealt{Dreizler20}) and TOI-132b (\citealt{Diaz20}). While the the exact mechanism responsible for clearing out this $R_p$-$P$ and $M_p$-$P$ region remains unknown, some models support a combination of photoevaporation and tidal disruption following high-eccentricity migration (\citealt{Mazeh16}; \citealt{Lundkvist16}; \citealt{Owen18}). 

Planets within the Hot Neptune Desert range from dense, atmosphere-poor mini-Neptunes to atmosphere-rich, "puffy" super-Neptunes. TOI-1272b lies in the middle of this spectrum, with an elevated density relative to the upper end of the \cite{Weiss14} relation. We used a 2-component composition model to determine the relative abundances of solid core and gaseous envelope for this dense Neptune, following the procedure of \cite{MacDougall21}. We interpolated over a 4D grid of stellar and planetary properties to derive the expected envelope mass fraction for TOI-1272b using the \cite{Lopez14a} planet structure models. Assuming an earth-like core composition and a solar-composition H/He envelope, we estimated $f_{\rm env} = 10.9 \pm 2.0\%$ and a core mass of $21.9 \pm 2.0$ $M_{\rm \earth}$. Given the strong stellar irradiance experienced by this planet, with $T_{\rm eq} \approx 960$ K (assuming albedo $\alpha = 0.3$), TOI-1272b could have begun as a more atmosphere-rich Neptune similar to GJ 3470b (\citealt{Kosiarek19}) and experienced subsequent atmosphere loss. TOI-1272b may then serve as a strong candidate for follow-up atmospheric observations, following the treatment of similar targets like those discussed by \cite{Crossfield17}.

The outer companion in this system, TOI-1272c, likely falls into the same size category as TOI-1272b, with $M_{\rm p,c}$ sin$i = 27.4 \pm 3.2$ $M_{\rm \earth}$. However, since no transit was detected in TESS photometry, we were unable to make any claims regarding its density or composition. One might suppose that a sufficiently low-radius planet on an 8.7-day orbit could produce a transit signal below the detection threshold of S/N $\approx$ 7.1. Assuming a transit duration of $T_{\rm 14} \approx 0.15$ days and the same noise properties as the TOI-1272b transit, this would require $R_{\rm p,c} \lesssim 2.3$ $R_{\rm \earth}$ and $\rho_{\rm p,c} \gtrsim 12.0$ g cm$^{-3}$. While this density is not entirely unreasonable (see, e.g., Kepler-411b; \citealt{Sun19}), it is unlikely given the known sample of similar planets.

\subsection{Eccentricities}
\label{sec:ecc}

\begin{figure}[ht]
\centering
\includegraphics[width=0.48
\textwidth]{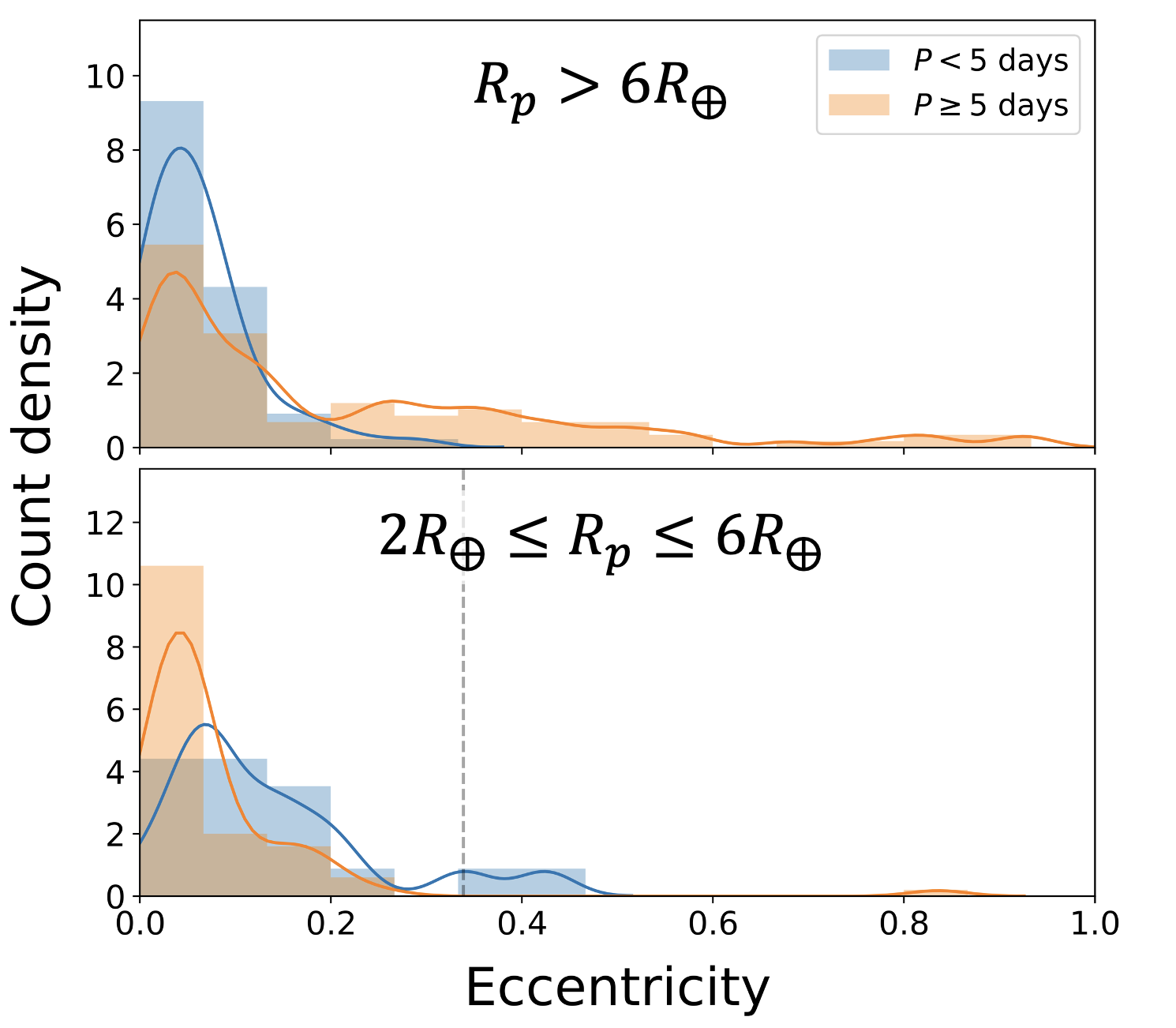}
\caption{Eccentricity distribution of planets with eccentricity uncertainties of $\sigma_e$ < 0.1, comparing shorter period planets (blue) against longer period planets (orange). The lines show kernel density estimation fits to the data of the corresponding color. Top: Jupiter-like planets ($>6$ $R_{\rm \earth}$). Bottom: Neptune-like planets (2--6 $R_{\rm \earth}$), including TOI-1272b (dashed gray line).}
\label{fig:ecc_distr}
\end{figure}

The eccentricity distribution of hot Neptunes was discussed in depth by \cite{Correia20} who noted that such planets exhibit elevated eccentricities despite being on compact orbits. We reconsidered this claim using a more recent set of confirmed planet data from the NASA Exoplanet Archive (\citealt{Akeson13}; \citealt{ps}), including TOI-1272b in our sample. We considered Neptunes to have radii $\sim$2--6 $R_{\rm \earth}$ and "hot" planets to have $P < 5$ days. Constraining our sample to only planets with eccentricity uncertainties less than 0.1 (Figure \ref{fig:ecc_distr}), we found that hot Neptunes ($N = 17$) displayed a broader eccentricity distribution than their longer period counterparts ($N = 75$). We verified the distinction between the two distributions using a Kolmogorov-Smirnov (KS) test, finding $p \approx 0.008$. TOI-1272b contributed to this significant trend among hot Neptunes.

On the other hand, planets with radii $>6$ $R_{\rm \earth}$ showed the opposite trend as can also be seen in Figure \ref{fig:ecc_distr}, verified by a KS test with $p \ll 0.01$. In this radius range, 66 planets had $P < 5$ d and 88 had $P \geq 5$ d. We did not consider planets with radii $<2$ $R_{\rm \earth}$ in this analysis due to the low sample size of such planets with eccentricity uncertainties $<0.1$. However, a cursory examination of this small subset suggested similar eccentricity distributions between shorter and longer period planets in this size range. 

While these findings are statistically significant based on KS tests given the current data, several confounding factors led us to determine that these eccentricity trends are suggestive rather than definitive at this time. These factors include the small sample size of hot Neptunes, the reliability of the data reported by the NASA Exoplanet Archive, and potential observational biases. Nonetheless, the possible disagreement between the eccentricity trend for hot Neptunes versus hot Jupiters is an active area of research. The low eccentricities of hot Jupiters were consistent with the rapid tidal circularization timescales expected at lower semi-major axes. Conversely, hot Neptunes seem more likely to violate the rule. 

\begin{figure}[ht]
\centering
\includegraphics[width=0.48
\textwidth]{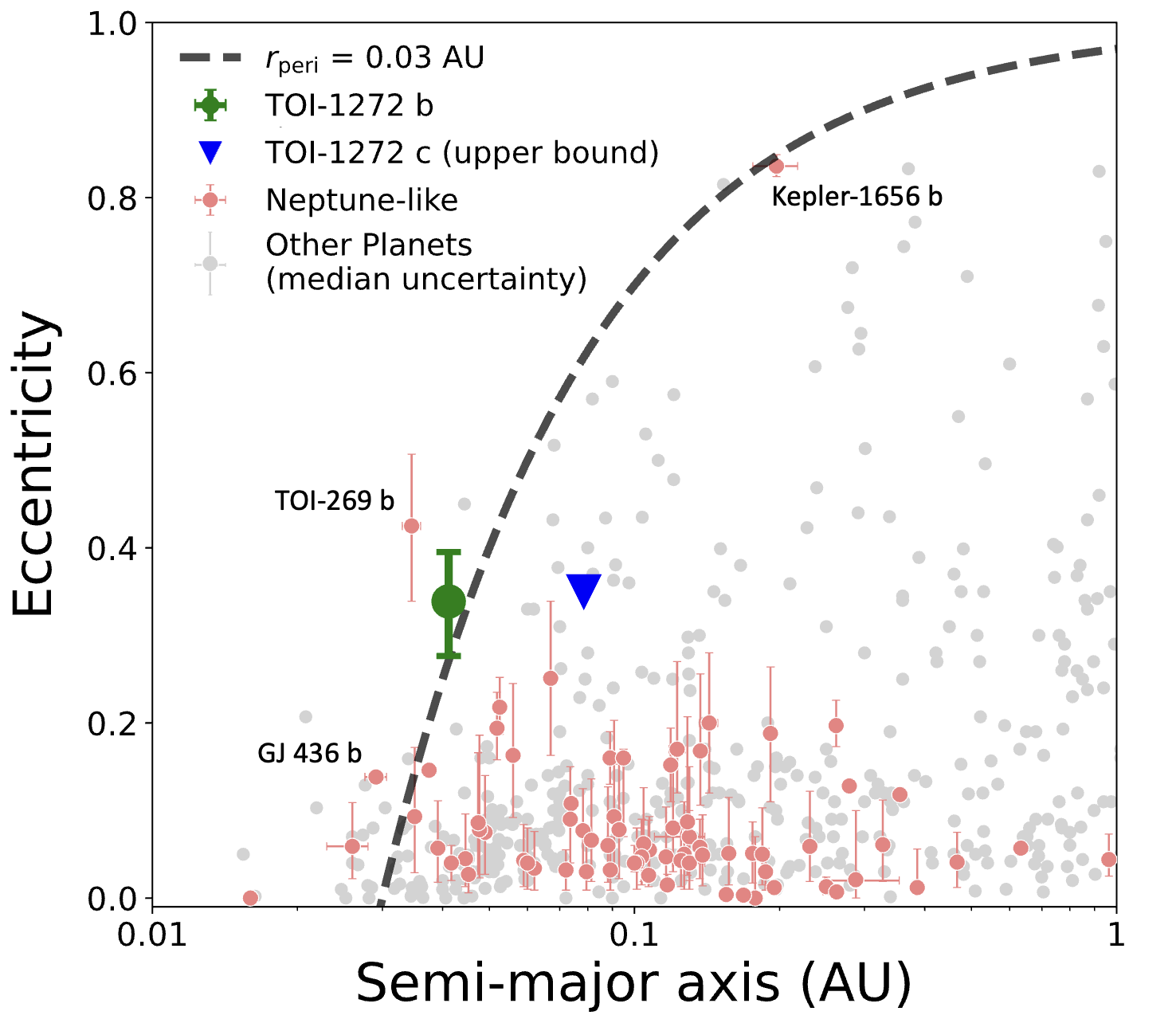}
\caption{Eccentricity distribution of planets with $\sigma_e$ < 0.1 (Neptune-like in red, other known planets in gray) as a function of orbital separation, showing TOI-1272 b and c in green and blue, respectively. Periastron distance of 0.03 AU is shown as an empirical threshold for rapid tidal eccentricity decay. Three eccentric Neptunes are labeled for reference (Kepler-1656b; TOI-269b, \citealt{Cointepas21}; GJ 436b).}
\label{fig:ecc_sep}
\end{figure}

A nominal empirically-derived periastron distance of $r_{\rm peri} \approx 0.03$ au is often used to approximate the boundary of rapid tidal eccentricity decay, shown in Figure \ref{fig:ecc_sep}. Here, we see that only a small subset of well-characterized planets inhabit the high-eccentricity area of parameter space beyond this boundary, including TOI-1272b and a few other eccentric hot Neptunes. The well-studied planet GJ 436b is among these Neptunes with orbits that disagree with tidal circularization, making it a near-twin to TOI-1272b based on mass, radius, eccentricity, and period. 

\subsection{A Unique Formation and Evolution Pathway}
\label{sec:formation}

The sparsity of the Hot Neptune Desert along with the counter-intuitive trend in hot Neptune eccentricities suggests a unique evolutionary pathway for hot planets within the Neptune size regime. Several studies have sought to explain the dearth of planets within the "desert" region of $M_p$-$P$ and $R_p$-$P$ parameter space. The leading hypothesis suggests a combination of photoevaporation (\citealt{Owen13}; \citealt{Owen18}) and high-eccentricity migration (\citealt{Mazeh16}; \citealt{Matsakos16}). Interestingly, photoevaporation is also cited as a possible mechanism for maintaining non-zero eccentricities among hot Neptunes (\citealt{Correia20}), along with planet-planet excitation (see, e.g. \citealt{Juric08}; \citealt{Chiang13}) or an Eccentric Kozai-Lidov (EKL) effect from a distant giant companion (\citealt{Naoz16}). The persisting eccentricities of some hot Neptunes could also simply be a result of $Q'$ values underestimated by an order of magnitude or more, which would make them inconsistent with the $Q'$ values measured for Neptune and Uranus through interior modeling.

An underestimated $Q'$ could certainly be the case for TOI-1272b, contributing to a longer $\tau_e$ and slower rate of eccentricity decay. TOI-1272b may also be undergoing significant photoevaporation given its $f_{\rm env}$ and close-in $r_{\rm peri}$, contributing to both its location in the middle of the Hot Neptune Desert and its high eccentricity. However, TOI-1272b differs from the plausible formation and evolution pathways of other hot Neptunes due to the presence of a stable, nearby outer companion. Both high-eccentricity migration and perturbations from a distant companion through EKL effects are complicated by the presence of the mildly eccentric companion on an 8.7-day orbit. Such excitation mechanisms would have likely caused an orbit-crossing event and subsequent ejection of one or both planets. 

Instead, we propose that, along with photoevaporation, TOI-1272b has experienced minor planet-planet excitation events with TOI-1272c, possibly involving close-approaches or resonance-crossing events during migration (\citealt{Ford08}). These events could have contributed to both the high eccentricity and possible inward migration of TOI-1272b into the Hot Neptune Desert region of parameter space, similar to the proposed evolution of mini-Neptune HIP-97166b (\citealt{MacDougall21}). The elevated eccentricity of the inner planet may then persist in spite of strong tidal forces through Laplace-Lagrange oscillations that continue to force the eccentricity, as seen in our dynamical simulations mentioned in \S\ref{sec:stability}. However, additional considerations such as the relative inclination of the two planets may be necessary for a more detailed description of the dynamical evolution of this system.

\section{Summary and Conclusions}
\label{sec:conclusions}

In this work, we introduced a newly discovered planet within the Hot Neptune Desert around TOI-1272 with a 3.3-day orbital period. We predicted that this planet might have a high eccentricity based on a mismatch between the observed transit duration and the expected duration for a circular orbit upon modeling 3 sectors of transit photometry from TESS. We confirmed this high eccentricity with follow-up RV measurements and verified its stability through dynamical constraints, yielding $e_b = 0.34 \pm 0.06$. We also identified a non-transiting outer companion on an 8.7-day orbit, placing a limit on its eccentricity of $e_c \lesssim 0.35$. TOI-1272b is now one of only a handful of close-in Neptunes with a well-constrained high eccentricity. The high eccentricity of this inner planet persists in spite of strong tidal forces, likely as a result of either underestimated tidal quality factors for close-in exo-Neptunes or stable dynamical interactions with the outer planet that continue to pump the eccentricity. Nonetheless, the discovery of TOI-1272 b and c has boosted the sample size of a small and poorly understood class of planets, contributing to ongoing studies of hot Neptunes and eccentric short-period planets.

\begin{acknowledgments}
The authors thank Daniel Foreman-Mackey for discussions regarding the parameterization of the transit models used in this work. The authors also thank Konstantin Batygin for discussions on the dynamical context of this planetary system. MM acknowledges support from the UCLA Cota-Robles Graduate Fellowship. 

DH acknowledges support from the Alfred P.\ Sloan Foundation, the National Aeronautics and Space Administration (80NSSC21K0652), and the National Science Foundation (AST-1717000). DD acknowledges support from the TESS Guest Investigator Program grant 80NSSC22K0185 and NASA Exoplanet Research Program grant 18-2XRP18\_2-0136. TF acknowledges support from the University of California President's Postdoctoral Fellowship Program. RAR is supported by the NSF Graduate Research Fellowship, grant No.\ DGE 1745301. PD is supported by a National Science Foundation (NSF) Astronomy and Astrophysics Postdoctoral Fellowship under award AST-1903811. JMAM is supported by the National Science Foundation Graduate Research Fellowship Program under Grant No. DGE-1842400. JMAM acknowledges the LSSTC Data Science Fellowship Program, which is funded by LSSTC, NSF Cybertraining Grant No.\ 1829740, the Brinson Foundation, and the Moore Foundation; his participation in the program has benefited this work. 

This work was supported by a NASA Keck PI Data Award, administered by the NASA Exoplanet Science Institute. Data presented herein were obtained at the W.\ M.\ Keck Observatory from telescope time allocated to the National Aeronautics and Space Administration through the agency's scientific partnership with the California Institute of Technology and the University of California. The Observatory was made possible by the generous financial support of the W.\ M.\ Keck Foundation.

We thank the time assignment committees of the University of California, the California Institute of Technology, NASA, and the University of Hawaii for supporting the TESS-Keck Survey with observing time at Keck Observatory.  We thank NASA for funding associated with our Key Strategic Mission Support project.  We gratefully acknowledge the efforts and dedication of the Keck Observatory staff for support of HIRES and remote observing.  We recognize and acknowledge the cultural role and reverence that the summit of Maunakea has within the indigenous Hawaiian community. We are deeply grateful to have the opportunity to conduct observations from this mountain.  

This paper is based on data collected by the TESS mission. Funding for the TESS mission is provided by the NASA Explorer Program. We also acknowledge the use of public TESS data from pipelines at the TESS Science Office and at the TESS Science Processing Operations Center. This article is also based on observations made with the MuSCAT2 instrument, developed by ABC, at Telescopio Carlos Sánchez operated on the island of Tenerife by the IAC in the Spanish Observatorio del Teide. This work is partly financed by the Spanish Ministry of Economics and Competitiveness through grants PGC2018-098153-B-C31. This work is also partly supported by JSPS KAKENHI Grant Number JP17H04574, JP18H05439, JP20J21872, JP20K14521, JP21K20376, JST CREST Grant Number JPMJCR1761, the Astrobiology Center of National Institutes of Natural Sciences (NINS) (Grant Number AB031010), and Grant-in-Aid for JSPS Fellows.

This paper also includes data that are publicly available from the Mikulski Archive for Space Telescopes (MAST). Resources supporting this work were provided by the NASA High-End Computing (HEC) Program through the NASA Advanced Supercomputing (NAS) Division at Ames Research Center for the production of the SPOC data products. This work also used computational and storage services associated with the Hoffman2 Shared Cluster provided by UCLA Institute for Digital Research and Education's Research Technology Group. 
\end{acknowledgments}

\facilities{TESS, Keck/HIRES, MuSCAT, MuSCAT2}

\software{We made use of the following publicly available Python modules: \texttt{exoplanet} (\citealt{Foreman-Mackey2021}), \texttt{PyMC3} (\citealt{pymc16}), \texttt{theano} (\citealt{theano16}), \texttt{LDTK} (\citealt{ldtk15}), \texttt{RadVel} (\citealt{Fulton18a}), \texttt{RVSearch} (\citealt{Rosenthal21}), \texttt{juliet} (\citealt{juliet}), \texttt{batman} (\citealt{batman}), \texttt{dynesty} (\citealt{dynesty}),
\texttt{astropy} (\citealt{astropy:2013}, \citealt{astropy:2018}), \texttt{isoclassify} (\citealt{Huber17}), \texttt{lightkurve} (\citealt{lightkurve18}), \texttt{matplotlib} (\citealt{Hunter07}), \texttt{numpy} (\citealt{harris2020array}), \texttt{scipy} (\citealt{SciPy20}), \texttt{rebound} (\citealt{Rein12}), \texttt{stardate} (\citealt{Angus19}), \texttt{TESS-SIP} (\citealt{Hedges20}), \texttt{SpecMatch-Synth} (\citealt{Petigura17b}), and \texttt{pandas} (\citealt{pandas}).}


\bibliographystyle{aasjournal}
\bibliography{adslib}

\clearpage

\end{document}